\documentclass[letterpaper,twocolumn,english,pre,final]{revtex4}
\usepackage[T1]{fontenc}
\usepackage[latin1]{inputenc}
\usepackage{color}
\usepackage{graphicx}

\makeatletter

\usepackage{psfrag}

\usepackage{babel}
\makeatother
\begin{document}

\title{Theory of microphase separation on Side-Chain Liquid-Crystalline
Polymers with flexible spacers}

\author{Marcela Hernández$^{1,2}$ and Harry Westfahl Jr.$^{2}$}

\affiliation{\textit{\small $^{1}$Instituto de Física {}``Gleb Wataghin'',
Universidade Estadual de Campinas, Caixa Postal 6165, Campinas SP
13083-970, Brazil }\\
\textit{\small $^{2}$Laboratório Nacional de Luz Síncrotron-ABTLuS,
Caixa Postal 6192, Campinas, SP 13043-090, Brazil}}

\begin{abstract}
We model a melt of monodisperse side-chain liquid crystalline polymers
as a melt of comb copolymers in which the side groups are rod-coil
diblock copolymers. We consider both excluded volume and Maier-Saupe
interactions. The first acts among any pair of segments while the
latter acts only between rods. Using a free energy functional calculated
from this microscopic model, we study the spinodal stability of the
isotropic phase against density and orientational fluctuations. The
phase diagram obtained in this way predicts nematic and smectic instabilities
as well as the existence of microphases or phases with modulated wave
vector but without nematic ordering. Such microphases are the result
of the competition between the incompatibility among the blocks and
the connectivity constrains imposed by the spacer and the backbone.
Also the effects of the polymerization degree and structural conformation
of the monomeric units on the phase behavior of the side-chain liquid
crystalline polymers are studied.
\end{abstract}
\maketitle

\section{Introduction}

\begin{figure}
\begin{center}\psfrag{ns}{$n_s$}

\psfrag{t1}{$\tau=1$}

\psfrag{j1}{$j=1$}

\psfrag{dt}{$\delta t$}

\psfrag{nr}{$n_r$}

\psfrag{jnt}{$j=n_t$}

\psfrag{tnb}{$\tau=n_b$}\includegraphics[%
  width=0.9\columnwidth,
  keepaspectratio]{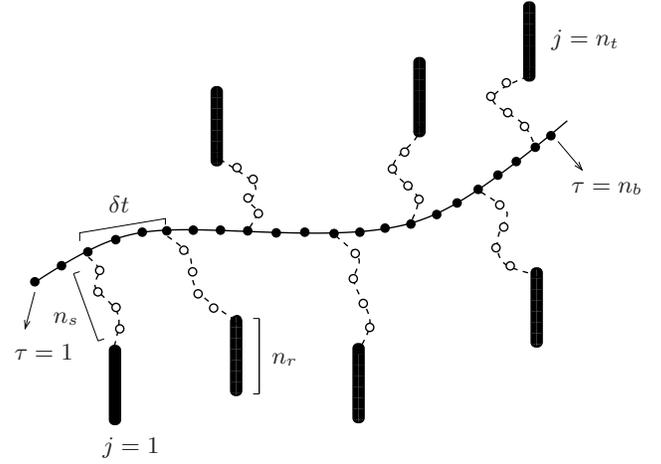}\end{center}

\caption{{\small \label{cap:Model-for-the} Model for the side chain polymer.
The backbone and the spacer are considered as Gaussian chains of $n_{b}$
and $n_{s}$ segments respectively while the rigid rods have $n_{r}$
segments. There are $n_{t}$ side groups separated by a distance $\delta t$
along the backbone.}}
\end{figure}

Side-chain liquid crystalline polymers (SCLCP) are comb copolymers
composed of a long main chain backbone with elongated rigid side chains
regularly attached through flexible spacers chain (schematic representation
on figure \ref{cap:Model-for-the}). Despite their high viscosity,
which can be two orders of magnitude higher than monomeric liquid
crystals, SCLCP are good candidates to storage information by locking
oriented structures into glassy states \cite{Shibaev1984}.

These interesting complexes can be though of as supramolecular polymers
of rod-coil diblock copolymers. In rod-coil diblock copolymers the
incompatibility between blocks competes against their connectivity
leading to a frustration of the macroscopic phase separation. This
competition, together with the tendency of the rod segments to align
with each other, results in a remarkably rich variety of microphase
geometries\cite{lee2001ssr,Yamazaki2004}. Such patterns are characterized
by alternating regions of each block with different collective orientations
of the rods\cite{Pryamitsyn2004}. The main difference between the
rod-coil diblock copolymers and the SCLCP is the introduction of additional
correlations among the rigid molecules brought by their linking through
the backbone\cite{Renz1986}\cite{finkelmann1991cel}. Based on the
accumulated theoretical and experimental knowledge on rod-coil diblock
copolymers\cite{Yamazaki2004}\cite{Reenders2002}\cite{Motoyama2003}\cite{matsen:4108}
and comb-like copolymers\cite{Shinozaki1994}\cite{Vlahos1987}\cite{Wang2005}
it is natural to expect that microphase segregation will also play
an important role in SCLCP.

The current experimental and theoretical understanding of the physical
properties of liquid crystalline polymers derives from the knowledge
on monomeric liquid crystals and their solutions with polymers. In
a pioneering work, Vasilenko, Shibaev and Khokhlov \cite{Sergei1985a}
developed a model for the SCLCP based on Matheson and Flory \cite{mathesonjr1981stm}
lattice theory for chains with rodlike sections. The emergence of
thermodynamic phases with complex nematic orderings was then described
as a purely entropic effect driven by the main chain flexibility.
This model has been later improved by Auriemma, Corradini and Vacatello\cite{AURIEMMA1990}
to better include steric repulsion effects. These theories neglect
the Maier-Saupe anisotropic interaction and should be thus valid in
cases where Maier-Saupe interaction between segments is negligible
when compared to steric repulsion. The study of anisotropic interactions
in SCLCP was first introduced by Wang and Warner \cite{Wang1987}.
Their theory describes the interplay between side-chains and backbone
nematic orderings by means of an extension of the Maier-Saupe theory.
In this case, the SCLCP melt is treated as a mixture of worm-like
molecules and rigid rods with their own nematic ordering tendencies
analogously to the model studied by ten Bosch, Maissa and Sixou \cite{Bosch1983}
for a solution of nematic polymers in liquid crystalline solvents.
The Maier-Saupe coupling (MS) between the backbone and the side-chains
is introduced through a constant that describes the effective flexibility
of the spacer and determines up to which extent the side and main
chains tend to be perpendicular. While in the case of a solution of
nematic polymers in liquid crystal solvents the MS coupling between
the backbone and the side chains is always positive, i.e., tends to
align them, in the SCLCP the coupling due to the effective spacer
attachment competes against the nematic coupling between the side-chain
and the backbone. This yields an effective nematic coupling between
the backbone and the side chains which can be either positive or negative,
depending on the relative strength of the two terms. In Wang and Warner
theory excluded volume interactions are not taken into account. Thus,
we should expect it to be valid in the limit where the Maier-Saupe
interaction is much stronger than excluded volume interactions. The
possibility of microphase separation driven from the competition between
the mesogens and the backbone + spacer complex is not considered in
the aforementioned theories.

In this paper, along the lines of Holyst and Schick\cite{Holyst1992}
and Reenders and ten Brinke\cite{Reenders2002}, we study the interplay
between microphase separation and nematic ordering in SCLCP. We treat
an intermediate region where both excluded volume effects and Maier-Saupe
interaction are present. We consider that the Maier-Saupe interaction
acts only between rods and that the excluded volume effect acts among
all monomers. This should be a reasonable approximation when both
spacer and backbone are flexible. We model the SCLCP as a comb polymer
composed of a backbone and spacer as in reference \cite{Shinozaki1994}
with rigid rods attached to the tips of the comb teeth. The comb polymer
melt is treated as a set of Gaussian chain with Flory-Huggins interaction
and Maier-Saupe interaction between rods. The Landau free energy functional
is derived and the spinodal stability curves are obtained as function
of the SCLCP structural and interaction parameters. Within our formulation
the main difference between a rod-coil polymer formed by the spacer
and rod and the SCLCP arises from the inter-rod and inter-spacer correlations
induced by the connectivity through the backbone. Since our main aim
in this paper is to study the microphase separation instabilities
we will not consider here the nematic ordering of the backbone.

\section{The model}

We model a melt of monodisperse $n_{p}$ side-chain polymers with
$N$ segments of statistical length $b$ as a chain of $n_{t}$ rod-coil
copolymers regularly attached to a backbone as schematically represented
on figure \ref{cap:Model-for-the}. The backbone and the spacer (i.e.,
the coil part of the rod-coil copolymer) are flexible Gaussian chains
of $n_{b}$ and $n_{s}$ monomers respectively. The rigid rods are
made of $n_{r}$ segments linearly arranged along a fixed direction.
It is useful to define the volume fraction of the rods, $f_{r}=\frac{n_{r}n_{t}}{N}$,
the volume fraction of the spacers $f_{s}=\frac{n_{s}n_{t}}{N}$ and
the volume fraction of the backbone segments , $f_{t}=\frac{n_{b}}{N}$. 

There are two important interactions that we consider in our model.
The first one is the steric repulsion between any monomers. The second
interaction is the Maier-Saupe potential\cite{maier1960sms}\cite{Stephen1974}
that tends to align the mesogens. We write the interaction Hamiltonian
as\begin{eqnarray}
\beta H_{I} & = & \sum_{K\neq K'\in(b,s,r)}\int d\mathbf{x}\,\epsilon_{KK'}\hat{\phi}_{K}\left(\mathbf{x}\right)\hat{\phi}_{K'}\left(\mathbf{x}\right)\nonumber \\
 &  & -\frac{\omega}{2}\int d\mathbf{x}\,\hat{Q}^{\mu\nu}\left(\mathbf{x}\right)\hat{Q}^{\nu\mu}\left(\mathbf{x}\right)\,,\label{eq:Hi}\end{eqnarray}
where $K$ and $K'$ represent the monomer species indexes, i.e, backbone
segment, spacer segment or rod segment, while $\epsilon_{KK'}$ and
$\omega$ are the strength of the interaction potentials. Equation
(\ref{eq:Hi}) is already written in terms of the volume fractions
$\hat{\phi}_{K}\left(\mathbf{x}\right)$ and the nematic tensor $\hat{Q}^{\mu\nu}\left(\mathbf{x}\right)$
of the rods, where the local volume fractions of the species are defined
as\begin{equation}
\hat{\phi}_{b}\left(\mathbf{x}\right)=v_{0}\sum_{\alpha=1}^{n_{p}}\int_{0}^{n_{b}}d\tau\,\delta\left[\mathbf{x}-\mathbf{X}_{\alpha}^{b}\left(\tau\right)\right],\label{eq:phib}\end{equation}

\begin{eqnarray}
\hat{\phi}_{s}\left(\mathbf{x}\right) & = & v_{0}\sum_{\alpha=1}^{n_{p}}\sum_{j=1}^{n_{t}}\int_{0}^{n_{s}}d\tau\,\delta\left[\mathbf{x}-\mathbf{X}_{\alpha}^{s,j}\left(\tau\right)\right.\nonumber \\
 &  & \left.-\mathbf{X}_{\alpha}^{b}\left(\left(j-\frac{1}{2}\right)\delta t\right)\right],\label{eq:phis}\end{eqnarray}

\begin{eqnarray}
\hat{\phi}_{r}\left(\mathbf{x}\right) & = & v_{0}\sum_{\alpha=1}^{n_{p}}\sum_{j=1}^{n_{t}}\int_{0}^{n_{r}}d\tau\,\delta\left[\mathbf{x}-\mathbf{u}_{\alpha}^{j}\,\tau\right.\nonumber \\
 &  & \left.-\mathbf{X}_{\alpha}^{s,j}\left(n_{s}\right)-\mathbf{X}_{\alpha}^{b}\left(\left(j-\frac{1}{2}\right)\delta t\right)\right],\label{eq:phir}\end{eqnarray}
where $v_{0}$ is the volume of each segment and $j$ labels the side-group
position along the backbone. For the backbone, $\mathbf{X}_{\alpha}^{b}\left(\tau\right)$
is the curve that describes the conformation of the chain $\alpha$
as a function of $\tau$, which labels the monomers in the chain.
For the spacers, separated by $\delta t=\frac{n_{b}}{n_{t}}$ segments,
$\mathbf{X}_{\alpha}^{s,j}\left(\tau\right)$ describes their conformation
starting at position $\mathbf{X}_{\alpha}^{b}\left(\left(j-\frac{1}{2}\right)\delta t\right)$
on the backbone. The unit vector $\mathbf{u}_{\alpha}^{j}$ defines
the orientation of the rod $j$, connected to the end of the spacer
at position $\mathbf{X}_{\alpha}^{s,j}\left(n_{s}\right)+\mathbf{X}_{\alpha}^{b}\left(\left(j-\frac{1}{2}\right)\delta t\right)$.

The nematic tensor of the rods is defined as 

\begin{eqnarray}
\hat{Q}^{\mu\nu}\left(\mathbf{x}\right) & = & v_{0}\sum_{\alpha=1}^{n_{p}}\sum_{j=1}^{n_{t}}\int_{0}^{n_{r}}d\tau\left(u_{\mu,\alpha}^{j}u_{\nu,\alpha}^{j}-\frac{1}{3}\delta_{\mu\nu}\right)\nonumber \\
 &  & \times\delta\left[\mathbf{x}-\mathbf{u}_{\alpha}^{j}\,\tau-\mathbf{X}_{\alpha}^{s,j}\left(n_{s}\right)\right.\nonumber \\
 &  & \left.-\mathbf{X}_{\alpha}^{b}\left(\left(j-\frac{1}{2}\right)\delta t\right)\right].\label{eq:Qmunu}\end{eqnarray}

For the sake of simplicity we will considered that the backbone and
the spacer are made of the same segments. In this way, they can described
by a single field $\hat{\phi}_{c}=\hat{\phi}_{b}+\hat{\phi}_{s}$. 

Following Gupta and Edwards \cite{Gupta1993} we rewrite the partition
function of the entire melt in terms of collective variables $\phi_{c},\phi_{r},\mathbf{Q}$
as\begin{equation}
\mathcal{Z}=\int\mathcal{D}\phi_{c}\int\mathcal{D}\phi_{r}\int\mathcal{D}\mathbf{Q}e^{-\beta H_{I}\left[\phi_{c},\phi_{r},\mathbf{Q}\right]}\mathcal{Z}_{0}\left[\phi_{c},\phi_{r},\mathbf{Q}\right]\label{eq:partition}\end{equation}
where $\mathcal{Z}_{0}$ is the partition function of independent
SCLCP molecules which yields the entropic contribution to the total
free energy and is given by

\begin{eqnarray}
\mathcal{Z}_{0} & \propto & \Pi_{\alpha=1}^{n_{p}}\Pi_{j=1}^{n_{t}}\int\mathcal{D}\mathbf{X}_{\alpha}^{b}\int\mathcal{D}\mathbf{X}_{\alpha}^{s,j}\int\mathcal{D}\mathbf{u}_{\alpha}^{j}\delta\left(\phi_{c}-\hat{\phi}_{c}\right)\nonumber \\
 &  & \times\delta\left(\phi_{r}-\hat{\phi}_{r}\right)\delta\left(\mathbf{Q-\hat{Q}}\right)\delta\left(\left|\mathbf{u}_{\alpha}^{j}\right|-1\right)\nonumber \\
 &  & \times e^{-\beta H_{0}\left[\mathbf{X}_{\alpha}^{b},\mathbf{X}_{\alpha}^{s,j}\right]}\,.\label{Z0}\end{eqnarray}

The Hamiltonian of unperturbed Gaussian chains, $H_{0}$, is written
as

\begin{eqnarray*}
\beta H_{0} & = & -\frac{3}{2Nb^{2}}\left[\int\mathcal{D}\tau\left(\frac{\partial\mathbf{X}_{\alpha}^{b}\left(\tau\right)}{\partial\tau}\right)^{2}\right.\\
 &  & \left.+\int\mathcal{D}\tau\left(\frac{\partial\mathbf{X}_{\alpha}^{s,\, j}\left(\tau\right)}{\partial\tau}\right)^{2}\right]\,.\end{eqnarray*}

As discussed by Shinozaki et al.\cite{Shinozaki1994}, this assumption
of Gaussian statistics for the chains is probably not the more realistic
since the chains might be stretched relative to the Gaussian case.
Nevertheless, for comb with an evenly spaced teeth and small monomer
density, the Gaussian description is still reasonable. 

We now use auxiliary fields $h_{c}$, $h_{r}$ and $h_{Q}^{\mu\nu}$
to express the Dirac deltas in equation (\ref{Z0}),

\begin{eqnarray}
\mathcal{Z}_{0}\left[\phi_{c},\phi_{r},\mathbf{Q}\right] & \propto & \int\mathcal{D}h_{c}\int\mathcal{D}h_{r}\int\mathcal{D}\mathbf{h}_{Q}e^{-i\int d\mathbf{r}\, h_{c}\phi_{c}}\nonumber \\
 &  & \times e^{-i\int d\mathbf{r}\, h_{r}\phi_{r}}e^{-i\int d\mathbf{r}\, h_{Q}^{\mu\nu}Q^{\mu\nu}}\nonumber \\
 &  & \times\exp\left\{ -F_{1}\left[h_{r},h_{c,}\mathbf{h}_{Q}\right]\right\} \,,\label{eq:Z0}\end{eqnarray}
where

\begin{eqnarray}
\exp\left\{ -F_{1}\left[h_{r},h_{c,}\mathbf{h}_{Q}\right]\right\}  & \equiv & \Pi_{\alpha=1}^{n_{p}}\Pi_{j=1}^{n_{t}}\int\mathcal{D}\mathbf{X}_{\alpha}^{b}\int\mathcal{D}\mathbf{X}_{\alpha}^{s,j}\nonumber \\
 &  & \times\int\mathcal{D}\mathbf{u}_{\alpha}^{j}\delta\left(\left|\mathbf{u}_{\alpha}^{j}\right|-1\right)\nonumber \\
 &  & \times e^{-\beta H_{0}\left[\mathbf{X}_{\alpha}^{b},\mathbf{X}_{\alpha}^{s,j}\right]}e^{i\int d\mathbf{r}\, h_{c}\hat{\phi}_{c}}\nonumber \\
 &  & \times e^{i\int d\mathbf{r}h_{r}\hat{\phi}_{r}}e^{i\int d\mathbf{r}h_{Q}^{\mu\nu}\hat{Q}^{\mu\nu}}\,.\label{eq:f1}\end{eqnarray}

Following references \cite{Gupta1993} and \cite{Liu1993}, the integrals
over the auxiliary fields in equation (\ref{eq:Z0}) are approximated
according to the steepest descent method. We write the entropic contribution
to the free energy, $F_{1}\left[h_{r},h_{c,}\mathbf{h}_{Q}\right]$,
as :

\begin{equation}
F_{1}\left[h_{r},h_{c,}\mathbf{h}_{Q}\right]=\left[h_{r}\, h_{c}\, h_{Q}^{\mu\nu}\right]G^{\left(2\right)\mu\nu\rho\sigma}\left[\begin{array}{c}
h_{r}\\
h_{c}\\
h_{Q}^{\rho\sigma}\end{array}\right]\,,\label{eq:f1expanded}\end{equation}
where $G^{\left(2\right)\mu\nu\rho\sigma}$ is the matrix of non interacting
pair correlation functions :

\begin{equation}
G^{\mu\nu\rho\sigma}=\left(\begin{array}{ccc}
G_{rr} & G_{rc} & G_{rQ}^{\rho\sigma}\\
G_{cr} & G_{cc} & G_{cQ}^{\rho\sigma}\\
G_{Qr}^{\mu\nu} & G_{Qc}^{\mu\nu} & G_{QQ}^{\mu\nu\rho\sigma}\end{array}\right)\,\,.\label{eq:correlation matrix.}\end{equation}

The non-interacting chains pair correlation functions are given by\[
G_{K\, K'}\left(\mathbf{x},\mathbf{x}'\right)\equiv\left\langle \hat{\phi}_{K}\left(\mathbf{x}\right)\hat{\phi}_{K'}\left(\mathbf{x}'\right)\right\rangle _{0}\]
\[
G_{K\, Q}^{\mu\nu}\left(\mathbf{x},\mathbf{x}'\right)\equiv\left\langle \hat{\phi}_{K}\left(\mathbf{x}\right)\hat{Q}^{\mu\nu}\left(\mathbf{x}'\right)\right\rangle _{0}\]
for $K,K'=r,c$, and\[
G_{Q\, Q}^{\mu\nu,\rho\sigma}\left(\mathbf{x},\mathbf{x}'\right)\equiv\left\langle \hat{Q}^{\mu\nu}\left(\mathbf{x}\right)\hat{Q}^{\rho\sigma}\left(\mathbf{x}'\right)\right\rangle _{0}\]
\textcolor{black}{where $\left\langle ...\right\rangle _{0}$ means
the average over the partition function of the unperturbed Gaussian
chains \ref{eq:Z0} and $\hat{\phi}_{K}\left(\mathbf{x}\right)$ and
$\hat{Q}^{\mu\nu}\left(\mathbf{x}\right)$ are defined in equations
(\ref{eq:phib}), (\ref{eq:phis}), (\ref{eq:phir}) and (\ref{eq:Qmunu}).}
We can now expand the total free energy of the system as a function
of the order parameters $\phi$ and $\mathbf{Q}$ following the same
procedure as in references \cite{Liu1993} and \cite{Reenders2002}.
At this point, we assume that the system is incompressible, i.e., 

\begin{equation}
\phi_{c}\left(\mathbf{x}\right)+\phi_{r}\left(\mathbf{x}\right)=1\,,\label{eq:incompress}\end{equation}
which for $\mathbf{q}\neq0$ means\[
\phi_{r,\mathbf{q}}=-\phi_{c,\mathbf{q}}\equiv\phi_{\mathbf{q}}\,.\]

After applying condition in Hamiltonian, the excluded volume interaction
term reduces to \[
\chi\int d\mathbf{x}\,\hat{\phi}\left(\mathbf{x}\right)^{2},\]
where $\chi=\epsilon_{cc}+\epsilon_{rr}-2\epsilon_{cr}$ is the usual
Flory- Huggins parameter\cite{teraoka2002ps}.

The second order expansion term, which defines the stability limits
of the isotropic phase against fluctuations of the order parameters,
is given by

\begin{equation}
\mathcal{F}\left[\phi_{\mathbf{q}},\mathbf{Q}_{\mathbf{q}}\right]=\frac{1}{2}\sum_{\mathbf{q}}\left[\begin{array}{cc}
\phi_{\mathbf{-q}} & ,Q_{\mathbf{-q}}^{\mu\nu}\end{array}\right]\Gamma_{\mathbf{q}}^{\left(2\right)\mu\nu\rho\sigma}\left[\begin{array}{c}
\phi_{\mathbf{q}}\\
Q_{\mathbf{q}}^{\rho\sigma}\end{array}\right].\label{eq:fspin}\end{equation}

The second order vertex $\Gamma^{\left(2\right)\mu\nu\rho\sigma}$
is the inverse of the matrix $G^{\mu\nu\rho\sigma}$, as calculated
in references \cite{Reenders2002} and \cite{Singh1994}, plus a diagonal
matrix that contains the terms for the Flory-Huggins interaction and
the Maier-Saupe potential:

\begin{equation}
\Gamma^{\left(2\right)\mu\nu\rho\sigma}=\left[G^{\mu\nu\rho\sigma}\right]^{-1}-\left(\begin{array}{cc}
2N\chi & 0\\
0 & \frac{2N\omega}{3}\end{array}\right).\label{eq:gamma2}\end{equation}

This approximation for the second order vertex, known as random phase
approximation (RPA) is considered as reasonable for dense mixtures
of strongly correlated polymer chains, where the fluctuations are
very small \cite{Vilgis2000}\cite{Gennes1979}. The calculation the
matrix (\ref{eq:gamma2}) follows along the same lines of reference
\cite{Reenders2002}. The major difference here is that there are
additional correlations between the rods and the spacers attached
to the same backbone, in contrast to the rod-coil copolymer studied
in reference \cite{Reenders2002}. For instance, the Fourier transform
of the rod-rod correlation function in the SCLCP is given by

\begin{eqnarray}
G_{rr}\left(\mathbf{q}\right) & = & f_{r}^{2}\left[\frac{1}{n_{t}}K_{rr}\left(lq\right)\right.\nonumber \\
 &  & \left.+e^{-2q^{2}R_{s}^{2}}F_{r}\left(lq\right)^{2}\left(D_{n_{t}}\left(qR_{b}\right)-\frac{1}{n_{t}}\right)\right]\label{eq:grr}\end{eqnarray}
where, $K_{rr}\left(x\right)=\frac{2}{x^{2}}\left[\cos\left(x\right)-1+x\, Si\left(x\right)\right]$,
is the rod-rod correlation function for the rod-coil copolymer studied
in reference \cite{Reenders2002}, $F_{r}\left(x\right)=\frac{Si\left(x\right)}{x}$
is the form factor of a rod of length $l$ and $D_{n}\left(x\right)=\frac{1}{2n^{2}}\frac{e^{-x^{2}}+\sinh\left(\frac{x^{2}}{n}\right)n-1}{\sinh^{2}\left(\frac{x^{2}}{2n}\right)}$
is the Debye function for the backbone. Here we use $l=n_{r}b$ as
the length of the rod and $R_{s}=\sqrt{\frac{n_{s}b^{2}}{6}}$ and
$R_{b}=\sqrt{\frac{n_{b}b^{2}}{6}}$ as the gyration radius of the
spacer and of the backbone respectively. 

Note that in the limit $n_{t}=1$ and $\delta t\to0$ the correlation
function (\ref{eq:grr}) reduces to the rod-rod correlation function
for the rod-coil copolymer studied in reference \cite{Reenders2002}.
In the opposite limit, for $n_{t}\gg1$, which is the case of SCLCP
the rod-rod correlation function is dominated by the inter-rod correlations
within a SCLCP and it is mostly influenced by the backbone and spacer
correlations. This is what leads to the main aspect of the physical
properties of SCLCP when compared to rod-coil copolymers and to the
strong dependence of the mesophases on the structural parameters of
the SCLCP \cite{finkelmann1991cel}\cite{Shibaev1984}. The other
correlations functions are shown in appendix \ref{sec:The-non-interacting}.

Also, as done in reference \cite{Reenders2002}, we consider a nematic
ordering parallel to the wave vector $\mathbf{q}$:\begin{equation}
Q^{\mu\nu}\left(\mathbf{q}\right)=Q\left(\mathbf{q}\right)\left(\frac{q^{\mu}q^{\nu}}{q^{2}}-\frac{\delta^{\mu\nu}}{3}\right).\label{eq:anzats}\end{equation}

\begin{figure}
\begin{center}\psfrag{30}{$30$}

\psfrag{25}{$25$}

\psfrag{20}{$20$}

\psfrag{15}{$15$}

\psfrag{10}{$10$}

\psfrag{5}{$5$}

\psfrag{0.3}{$0.3$}

\psfrag{0.4}{$0.4$}

\psfrag{0.5}{$0.5$}

\psfrag{0.6}{$0.6$}

\psfrag{0.7}{$0.7$}

\psfrag{0.8}{$0.8$}

\psfrag{cnm}{$\chi \frac{N}{n_t}$}

\psfrag{nm=25}{$\frac{N}{n_t} =25$}

\psfrag{w/c=1}{$\frac{\omega}{\chi}=1$}

\psfrag{dt=1n=1}{$\delta t=1, n_t=1$}

\psfrag{dt=3n=5}{$\delta t=3, n_t=5$}

\psfrag{dt=3n=30}{$\delta t=3, n_t=15$}

\psfrag{dt=3n=15}{$\delta t=3, n_t=30$}

\psfrag{dt=10n=30}{$\delta t=10, n_t=1$}

\psfrag{fr}{$f_r$}

\psfrag{a}{$(a)$}

\psfrag{IN}{$IN curve$}

\psfrag{IM}{$IM curves$}\includegraphics[%
  width=0.9\columnwidth,
  keepaspectratio]{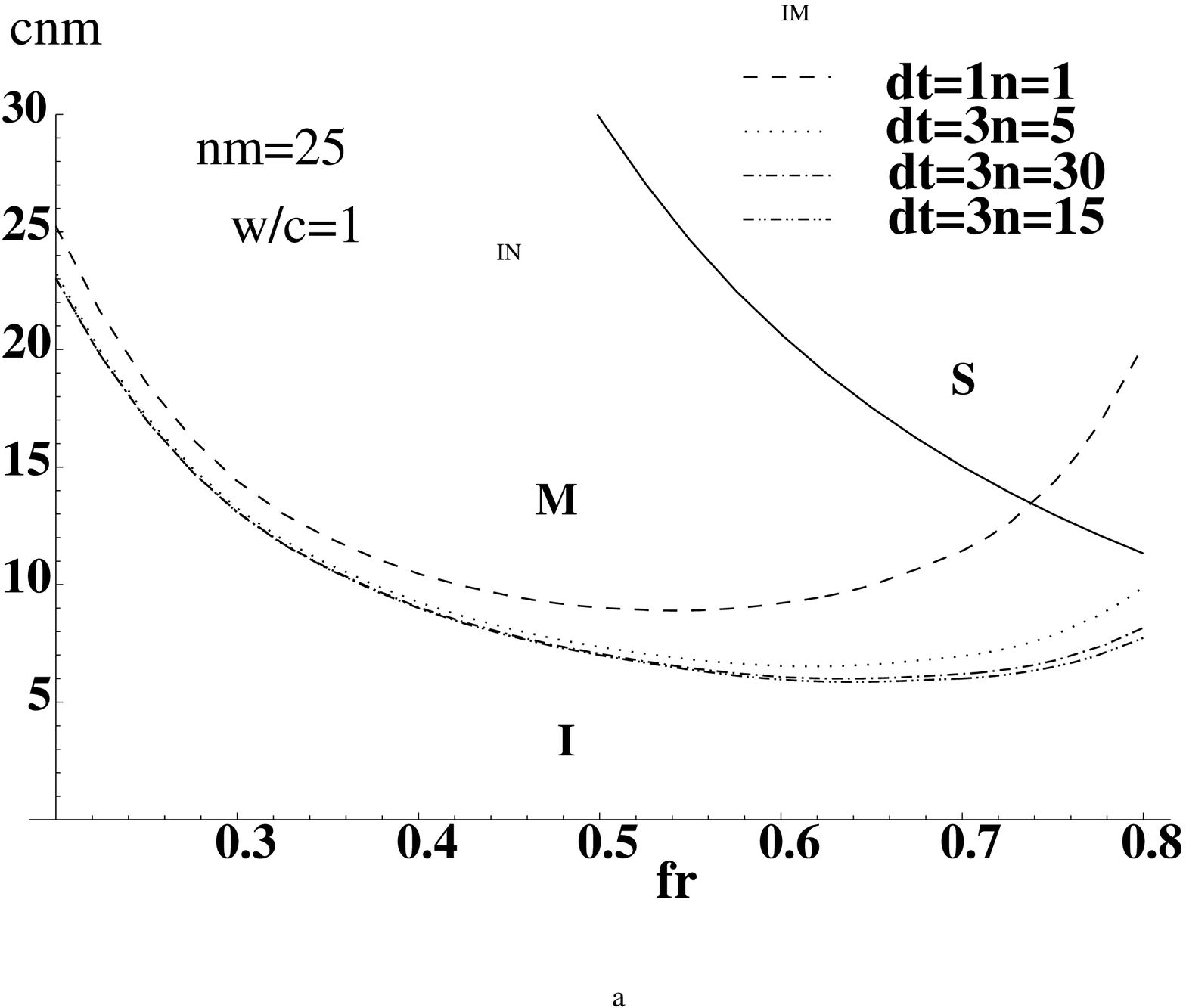}\end{center}

\psfrag{30}{$30$}

\psfrag{25}{$25$}

\psfrag{20}{$20$}

\psfrag{15}{$15$}

\psfrag{10}{$10$}

\psfrag{5}{$5$}

\psfrag{0.3}{$0.3$}

\psfrag{0.4}{$0.4$}

\psfrag{0.5}{$0.5$}

\psfrag{0.6}{$0.6$}

\psfrag{0.7}{$0.7$}

\psfrag{0.8}{$0.8$}

\psfrag{cnm}{$\chi \frac{N}{n_t}$}

\psfrag{n=25}{$\frac{N}{n_t} =25$}

\psfrag{w/c=8}{$\frac{\omega}{\chi}=8$}

\psfrag{t=1n=1}{$\delta t=1, n_t=1$}

\psfrag{t=3n=5}{$\delta t=3, n_t=5$}

\psfrag{t=3n=30}{$\delta t=3, n_t=15$}

\psfrag{t=3n=15}{$\delta t=3, n_t=30$}

\psfrag{t=10n=30}{$\delta t=10, n_t=1$}

\psfrag{fr}{$f_r$}

\psfrag{a}{$(b)$}

\psfrag{IN}{$IN curve$}

\psfrag{IM}{$IM curves$}\includegraphics[%
  width=0.9\columnwidth,
  keepaspectratio]{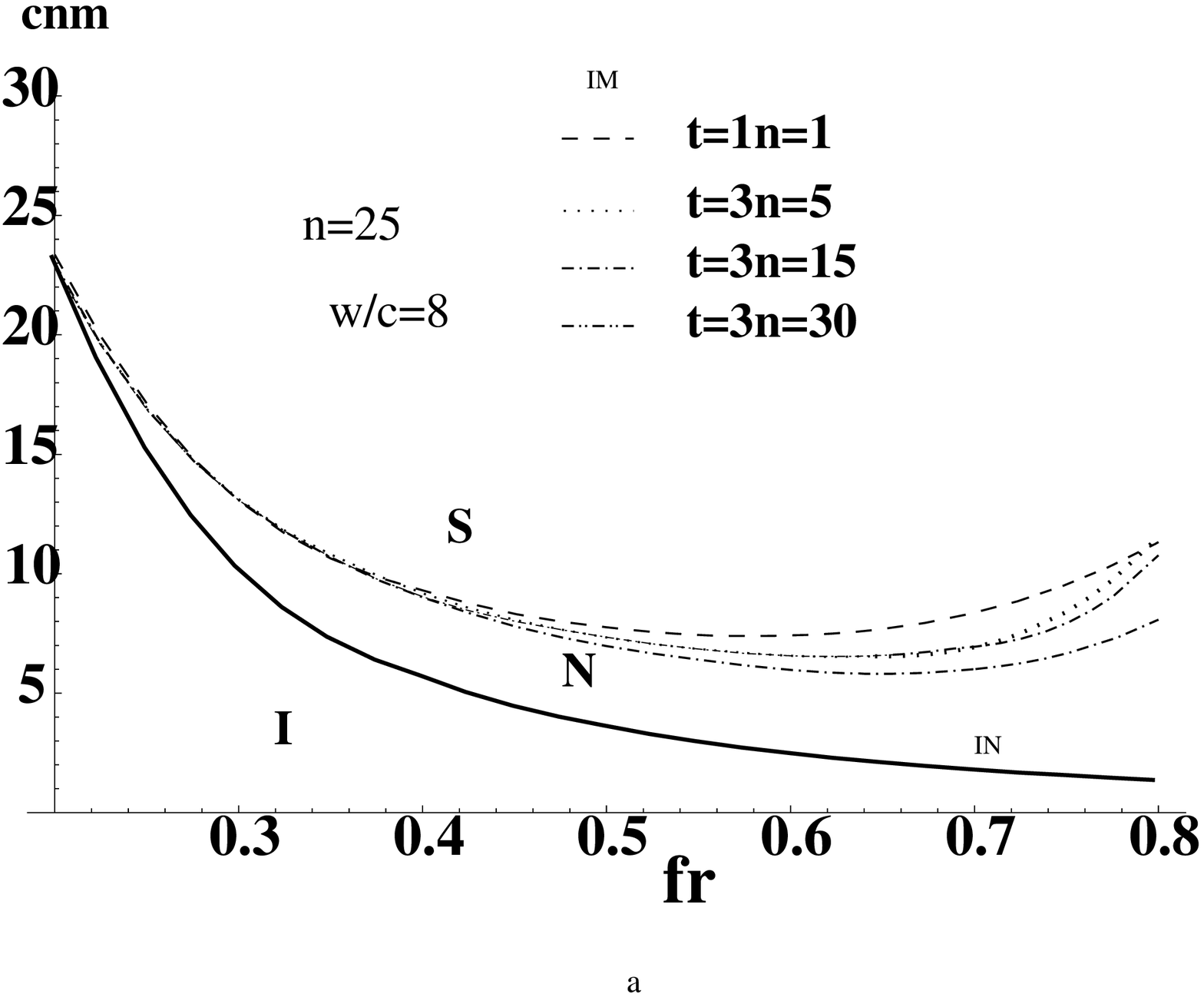}

\caption{\label{cap:Rod-coil-phase-diagram} Effect of the polymerization
degree and the relation between the Maier-Saupe and Flory-Huggins
interactions on the phase diagram for SCLCP for (a) $\frac{\omega}{\chi}=1$
(b) $\frac{\omega}{\chi}=8$. The isotropic phase is denoted by {}``I'',
while the microphases, nematic and smectic phases are labeled as {}``M'',
{}``N'' and {}``S'' respectively. The diagrams were calculated
for $\frac{N}{n_{t}}=25$, different values of and $n_{t}$ and $\delta t$
as indicated .}
\end{figure}

\section{Phase behavior}

\subsection{Stability limits of the isotropic phase: phase diagram}

The stability of the isotropic phase against fluctuations of the order
parameters is given by the spinodal curve. The roots of the determinant
of $\Gamma^{\left(2\right)\mu\nu\rho\sigma}$ define an expression
for $\chi$ as a function of the wave vector $q$, the structural
parameters $n_{r}$, $n_{s}$, $n_{t}$, $\delta t$ and the ratio
between Maier-Saupe and Flory Huggins interactions: 

\begin{equation}
\textrm{det}\left[\Gamma^{\left(2\right)\mu\nu\rho\sigma}\right]=0\rightarrow\chi\left(q,n_{r},n_{s},n_{t},\delta t,\frac{\omega}{\chi}\right).\label{eq:spincurve}\end{equation}

For $n_{r}$, $n_{s}$, $n_{t}$, $\delta t$ and $\frac{\omega}{\chi}$
fixed, the spinodal curve is determined by the minimization of $\chi$
with respect to $q$. This has been calculated using a program written
for Mathematica and available from the authors by request. As discussed
in reference \cite{Singh1994}, if the value $q^{*}$ that minimizes
equation (\ref{eq:spincurve}) is equal to zero, the isotropic phase
is unstable to a nematic phase. If $q^{*}\neq0$, then the isotropic
phase is unstable against modulations of the volume fraction, which
gives raise to microphase separation. On that basis, it is possible
to construct a phase diagram that sketches the the transition from
an isotropic phase to an ordered stated as a function of the structural
parameters of the model. For shortness and simplicity, in this paper
we denote as IN the spinodal obtained from the condition $q^{*}=0$
and IM the spinodal calculated according to $q^{*}\neq0$.

\subsection{Effect of the polymerization degree}

In figure \ref{cap:Rod-coil-phase-diagram}, we show the effect of
the polymerization degree on the phase diagram for two different values
of the ratio $\frac{\omega}{\chi}$. In both cases, the phase diagram
for $\delta t=1$, $n_{t}=1$ is the same as the rod-coil diblock
copolymer phase diagram calculated in reference \cite{Reenders2002}.
The region under the curves corresponds to an isotropic phase, denoted
as {}``I'' in the figure. Above them, the phase can be either nematic
(N), microphase (M) or smectic (S), which is superposition of the
nematic ordering and density modulation. When the Maier-Saupe interaction
is of the same order of the Flory-Huggins interaction, i.e. $\frac{\omega}{\chi}=1$,
the phase diagram shows the possibility of having microphase segregation
before the transition to a smectic state, like in phase diagram (a)
of figure \ref{cap:Rod-coil-phase-diagram}. If the value of the ratio
$\frac{\omega}{\chi}$ is such that there is no crossing point between
the two spinodals, as in phase diagram (b) of figure \ref{cap:Rod-coil-phase-diagram},
there is no microsegregation for any volume fraction. This might be
attributed to the fact that in this case, the Maier-Saupe interaction
is stronger than the Flory-Huggins interaction. Therefore therefore
the nematic ordering occurs first than the modulation of the density.

We note that the spinodal curves for the isotropic-nematic instabilities,
are given by\begin{equation}
\omega=\frac{15}{2f_{r}^{2}\left(\frac{N}{n_{t}}\right)}\,.\label{eq:MS}\end{equation}

This instability line does not depend on the polymerization degree
of the SCLCP but only on the ratio $\frac{N}{n_{t}}$, i.e, the total
number of monomers in each rod-coil comb plus $\delta t$. This is
exactly the same dependence as the case of a single rod-coil studied
in reference \cite{Reenders2002}. It is a consequence of the approximated
model we adopted where the rods only interact through the Maier-Saupe
coupling among themselves. Since the isotropic-nematic instability
occurs at the wave-vector $q=0$, the inter-rod coupling along a SCLCP
chain is not influenced by the backbone or the spacer.

The spinodal curves for the isotropic-microphase instabilities are
shifted towards higher temperatures as the degree of polymerization
increases. According to Finkelmann ad Rehage \cite{Finkelmann1984},
this behavior is well established experimentally, and can be understood
by the restriction of the translational and rotational motions of
the mesogenic molecules due to the linkage to the backbone. After
$n_{t}\approx10$, the chain length is such that there is no correlation
between the first monomer in the chain and the last added. Therefore,
the instability curve is hardly affected by the polymerization degree. 

\begin{figure}
\psfrag{0.8}{$0.8$}

\psfrag{0.6}{$0.6$}

\psfrag{0.4}{$0.4$}

\psfrag{0.3}{$0.3$}

\psfrag{0.2}{$0.2$}

\psfrag{0.3}{$0.3$}

\psfrag{0.4}{$0.4$}

\psfrag{0.5}{$0.5$}

\psfrag{0.6}{$0.6$}

\psfrag{0.7}{$0.7$}

\psfrag{0.8}{$0.8$}

\psfrag{fr}{$f_r$}

\psfrag{qstar}{$q^*$}

\psfrag{qrs}{$\frac{2 \pi}{n_r+R_s}$}

\psfrag{nt=1}{$n_t=1$}

\psfrag{nt=5}{$n_t=5$}

\psfrag{nt=15}{$n_t=15$}

\includegraphics[%
  width=0.9\columnwidth]{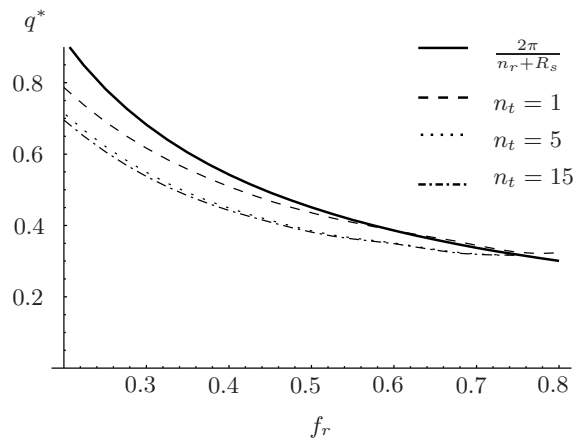}

\caption{\label{cap:qstar}Value of the wave vector $q*$ that defines spinodal
for the formation of microphases.The structural parameters were set
as $n_{t}=30$, $\delta t=3,$ and $n_{m}=25$. }
\end{figure}

The spinodal curve IM is also influenced by the ratio $\frac{\omega}{\chi}$
, specially in the rod-rich regions, as can be seen comparing phase
diagrams (a) and (b) of figure \ref{cap:Rod-coil-phase-diagram}.
However, this dependence vanishes with increasing $n_{t}$ and after
$n_{t}=5$ the spinodal curve IM is not longer affected by the ratio
$\frac{\omega}{\chi}$.

\subsection{Characteristic length}

In figure \ref{cap:qstar}, we show the value of $q^{*}$ that defines
the spinodal curve IM as a function of the rod volume fraction. The
solid line is the wave vector for the length of the rod-coil part
of the SCLCP, i.e. $q_{rs}=\frac{2\pi}{n_{r}+\sqrt{\frac{n_{s}}{6}}}$.
As it can bee seen from the figure, even in the rod-rich region, the
modulation period of the microphase is determined by both rod length
and spacer conformation. Nevertheless, up to $f_{r}\approx0.8$, $q^{*}$
is always smaller than $q_{rs}$ . As the polymerization degree is
increased (see dotted and dashed lines in figure \ref{cap:qstar}),
this difference increases. On the basis of this information and the
experimental data on SCLCP discussed by Noirez and coworkers in references
\cite{Noirez1998}\cite{noirez1995sac}\cite{noirez1994scl}, we suggest
the microphase region is formed by structures in which the backbone
is confined between layers containing the rigid cores. Therefore,
the difference between $q^{*}$ and $q_{rs}$ might be attributed
to the conformation of the backbone.

\subsection{Temperature equivalence}

So far, all the information we have discussed is based on the calculation
of the Flory-Huggins parameters as a function of the structural parameters
of the model and the ratio $\frac{\omega}{\chi}$. Since it is well
known that the Flory-Huggins parameter has a dependence on the inverse
of temperature of the form

\begin{equation}
\chi=\frac{A}{T}+B,\label{eq:chiflory}\end{equation}
it is interesting to relate the constants $A$ and $B$ to the SCLCP
chemical constitution in order to give a more realistic characteristic
to our results.

From thermodynamic arguments applied on the Flory's lattice model
for a polymer in a solvent, it can be shown that the temperature dependent
term of expression (\ref{eq:chiflory}) is the enthalpic contribution
from rearrangement of the contacts polymer-polymer, solvent-polymer
and solvent-solvent . See for example reference \cite{teraoka2002ps}
for a review on this subject. This term can be estimated using the
Hildebrand solubility parameters of the polymer and the solvent segments.
Following \cite{grulke1998ph}, the enthalpic contribution to the
Flory-Huggins parameter is 

\begin{equation}
A=\frac{V_{seg}}{R}\left(\delta_{pol}-\delta_{solv}\right)^{2},\label{eq:chi-hildebrand}\end{equation}
where the $\delta's$ are the Hildebrand solubility parameters of
each component of the mixture, $V_{seg}$ is the common segment volume
and $R$ is the ideal gas constant. The Hildebrand solubility parameters
are approximated through the relationship with the cohesive energy
and the group contribution method \cite{grulke1998ph} .

The athermal term in expression (\ref{eq:chiflory}) is related to
internal degrees of entropy of the polymers. According to different
authors it might have its origin in features such as conformational
asymmetries of the components of the mixture \cite{almdal2002ica,fredrickson1994ecf}
and nematic interaction between the segments originated by the flexibility
differences between the polymers. In general, this term is determined
empirically\cite{liu1992inf}. 

\begin{figure}
\begin{center}\psfrag{400}{$400$}

\psfrag{350}{$350$}

\psfrag{300}{$300$}

\psfrag{250}{$250$}

\psfrag{200}{$200$}

\psfrag{150}{$150$}

\psfrag{5}{$5$}

\psfrag{10}{$10$}

\psfrag{15}{$15$}

\psfrag{20}{$20$}

\psfrag{T}{$T \left( K \right)$}

\psfrag{n}{$n_s$}

\psfrag{w/c=3}{$\frac{\omega}{\chi}=3$}

\psfrag{I}{$I$}

\psfrag{M}{$M$}

\psfrag{N}{$N$}

\psfrag{S}{$S$}

\psfrag{a}{$(a)$}

\psfrag{IN}{$IN curve$}

\psfrag{IM}{$IM curve$}\includegraphics[%
  width=0.9\columnwidth]{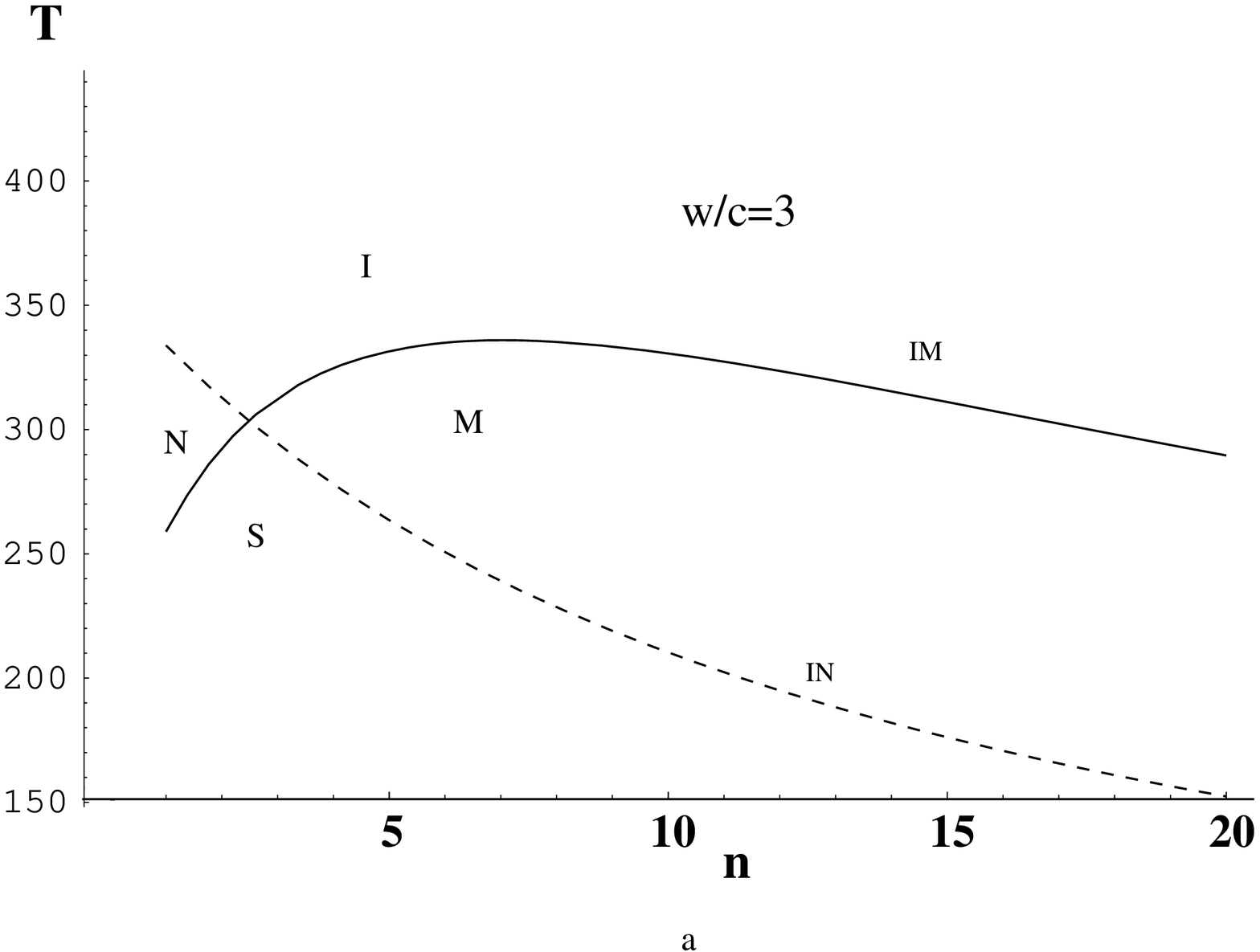}\end{center}

\psfrag{400}{$400$}

\psfrag{350}{$350$}

\psfrag{300}{$300$}

\psfrag{250}{$250$}

\psfrag{200}{$200$}

\psfrag{150}{$150$}

\psfrag{5}{$5$}

\psfrag{10}{$10$}

\psfrag{15}{$15$}

\psfrag{20}{$20$}

\psfrag{T}{$T \left( K \right) $}

\psfrag{n}{$n_s$}

\psfrag{w/c=5}{$\frac{\omega}{\chi}=5$}

\psfrag{I}{$I$}

\psfrag{M}{$M$}

\psfrag{N}{$N$}

\psfrag{S}{$S$}

\psfrag{a}{$(b)$}

\psfrag{IN}{$IN curve$}

\psfrag{IM}{$IM curve$}\includegraphics[%
  width=0.9\columnwidth]{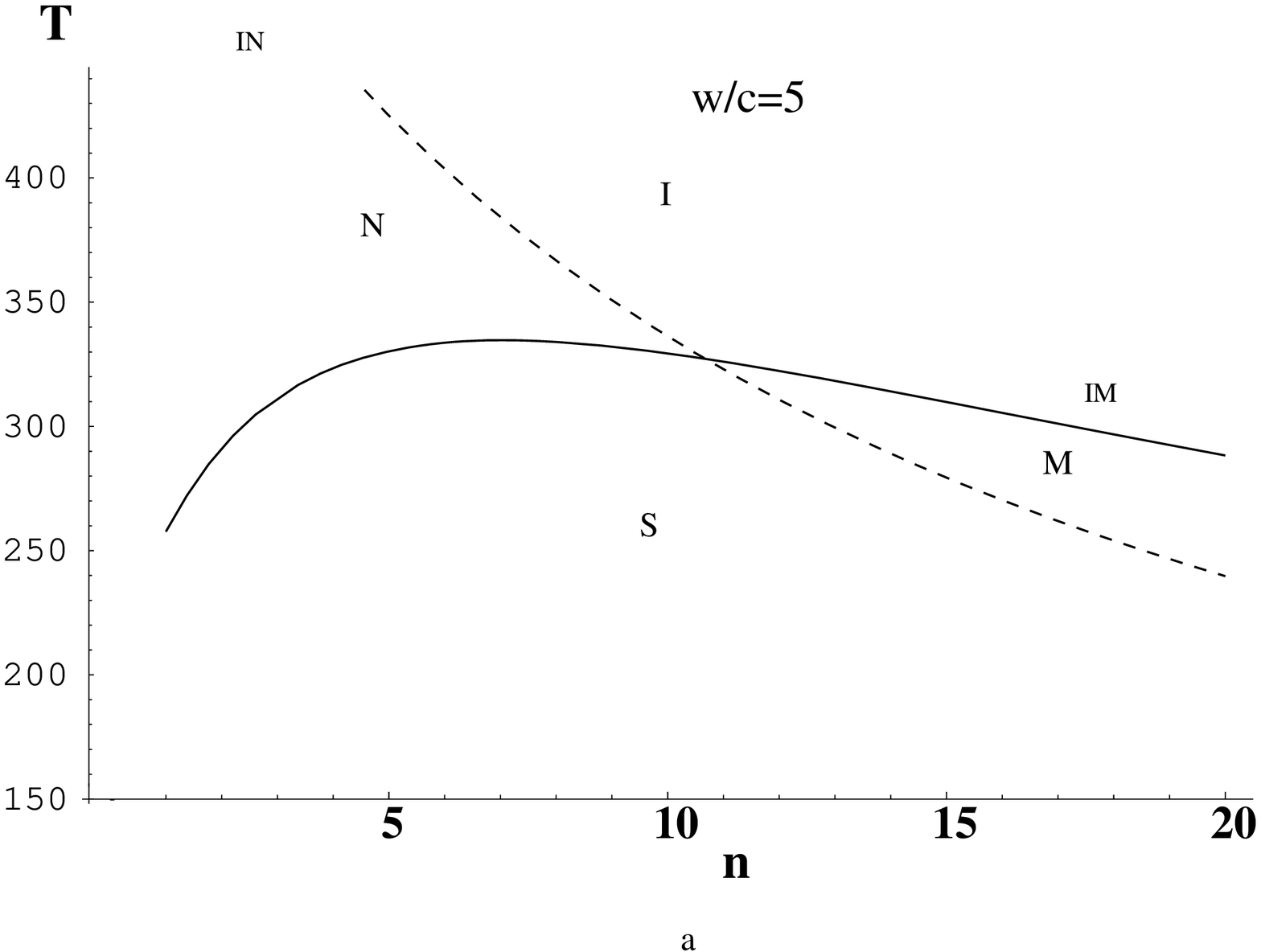}

\caption{.\label{cap:Transition-temperatures-as}Transition temperatures as
a function of the spacer length.Parameters:$\delta t=3$, $n_{r}=10$,
$n_{t}=30$}
\end{figure}

\begin{figure}
\begin{center}\psfrag{400}{$400$}

\psfrag{350}{$350$}

\psfrag{300}{$300$}

\psfrag{250}{$250$}

\psfrag{200}{$200$}

\psfrag{150}{$150$}

\psfrag{2}{$2$}

\psfrag{4}{$4$}

\psfrag{6}{$6$}

\psfrag{8}{$8$}

\psfrag{10}{$10$}

\psfrag{12}{$12$}

\psfrag{14}{$14$}

\psfrag{Tcrit}{$T \left( K \right)$}

\psfrag{ns}{$n_s$}

\psfrag{w/c=3}{$\frac{\omega}{\chi}=3$}

\psfrag{I}{$I$}

\psfrag{M}{$M$}

\psfrag{N}{$N$}

\psfrag{S}{$S$}

\psfrag{a}{$(a)$}

\psfrag{IN}{$IN curve$}

\psfrag{IM}{$IM curve$}\includegraphics[%
  width=0.9\columnwidth]{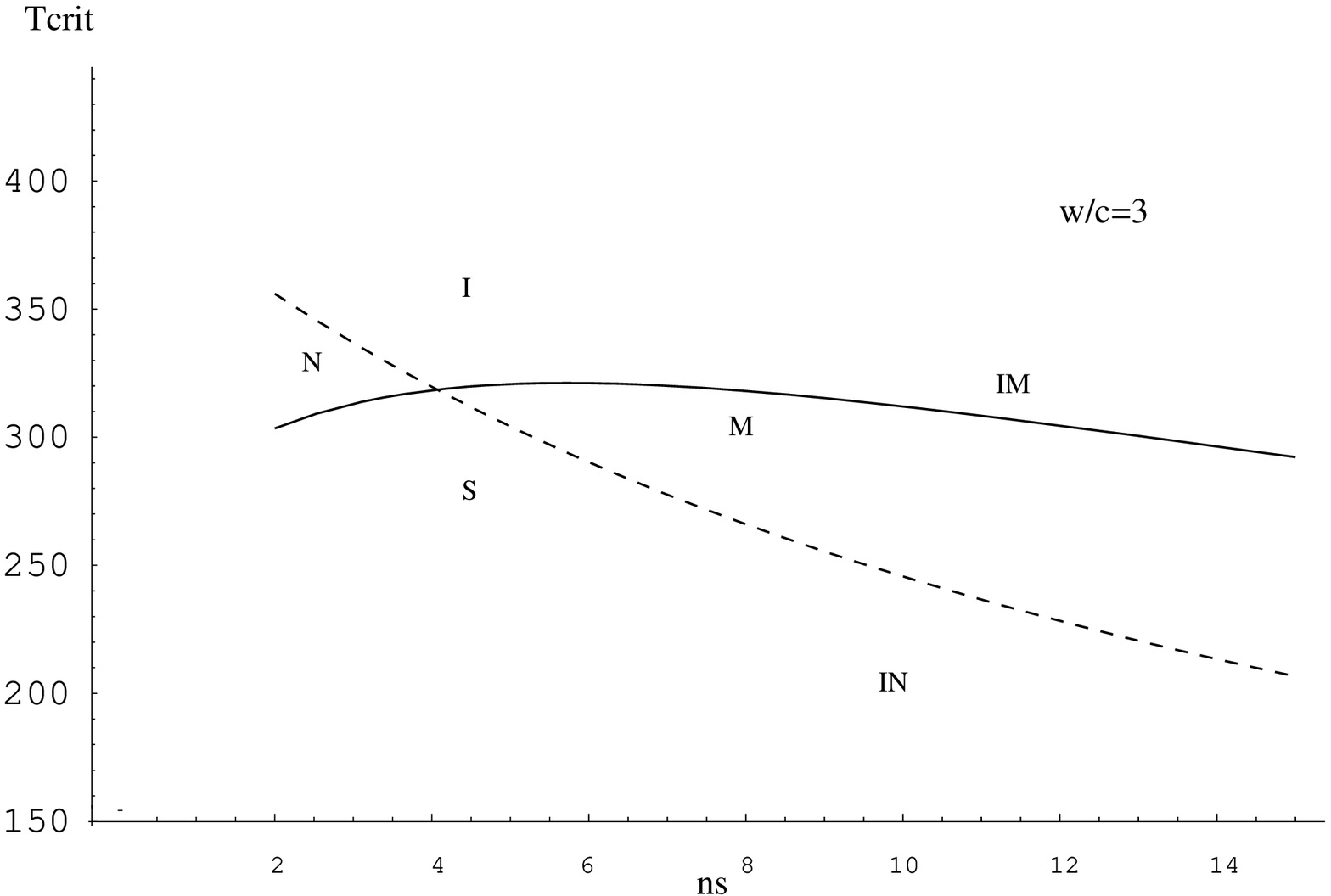}\end{center}

\psfrag{400}{$400$}

\psfrag{350}{$350$}

\psfrag{300}{$300$}

\psfrag{250}{$250$}

\psfrag{200}{$200$}

\psfrag{150}{$150$}

\psfrag{2}{$2$}

\psfrag{4}{$4$}

\psfrag{6}{$6$}

\psfrag{8}{$8$}

\psfrag{10}{$10$}

\psfrag{12}{$12$}

\psfrag{14}{$14$}

\psfrag{Tcrit}{$T\left( K \right)$}

\psfrag{ns}{$n_s$}

\psfrag{w/c=5}{$\frac{\omega}{\chi}=5$}

\psfrag{I}{$I$}

\psfrag{M}{$M$}

\psfrag{N}{$N$}

\psfrag{S}{$S$}

\psfrag{a}{$(b)$}

\psfrag{IN}{$IN curve$}

\psfrag{IM}{$IM curve$}\includegraphics[%
  width=0.9\columnwidth]{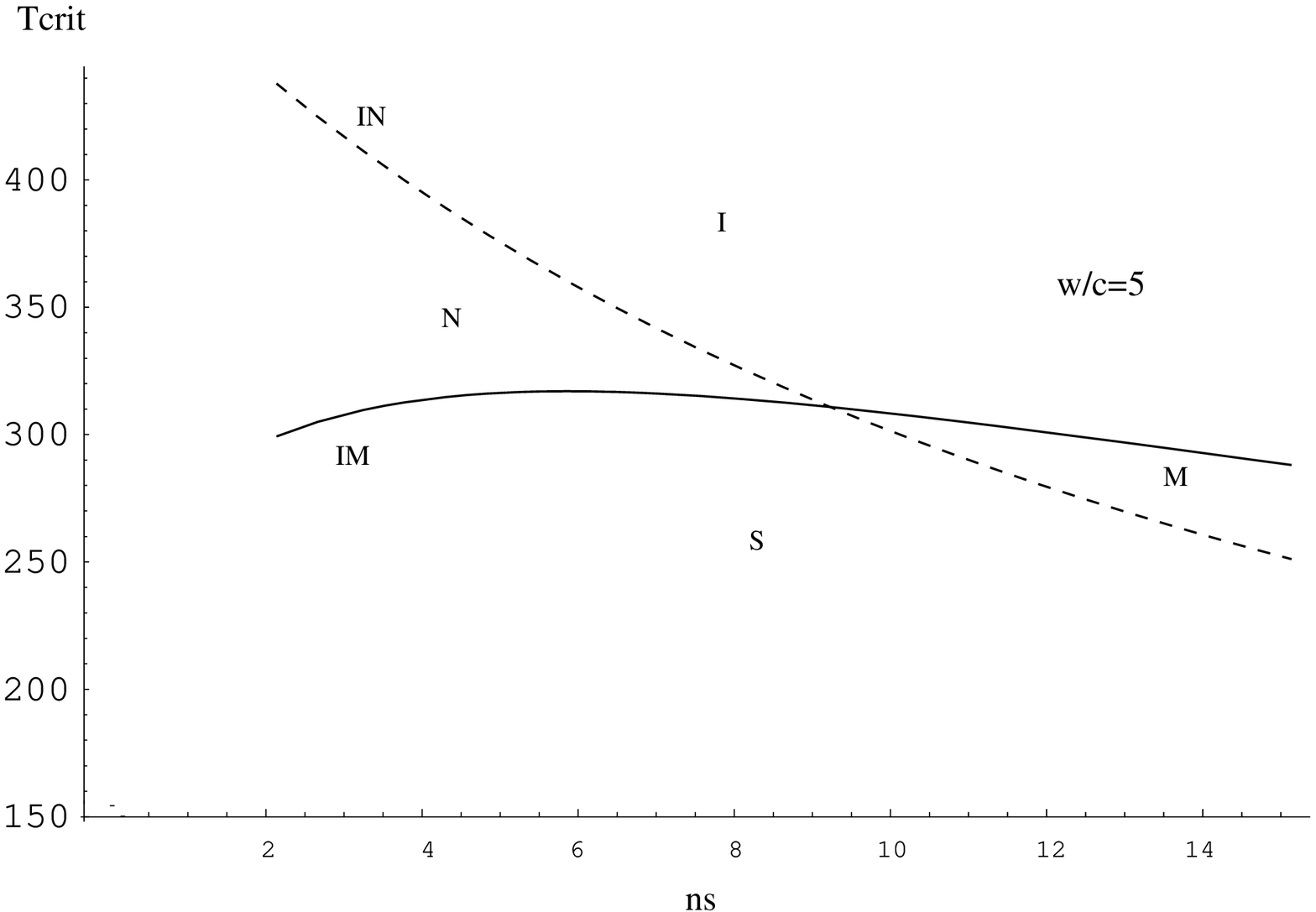}

\caption{Transition temperatures as a function of the spacer length.Parameters:$\delta t=5$,
$n_{r}=10$, $n_{t}=30$.\label{cap:Transition-temperatures-dt5}}
\end{figure}

\begin{figure}
\psfrag{CH}{$CH_3$}

\psfrag{C}{$C$}

\psfrag{O}{$O$}

\psfrag{CH2}{$\left( CH_2 \right)_{\delta t}$}

\psfrag{(CH2)}{$\left ( CH_2 \right )_{n_s}$}

\psfrag{OCH3}{$OCH_3$}

\psfrag{nrb}{$n_rb$}

\psfrag{n}{$n_t$}

\includegraphics[%
  width=0.9\columnwidth]{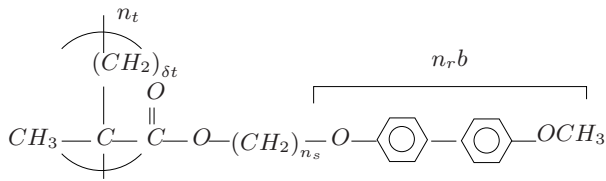}

\caption{\label{cap:Polycarylate-used-as}Polyacrylate used as example to
calculate the dependence of $\chi$ on temperature.}
\end{figure}

This equivalence between temperature and the $\chi$ parameter allows
us to compare our results with experimental data available in literature.
As an example, in appendix \ref{sec:The-Hildebrand-parameters} we
calculated the Flory-Huggins parameter as a function of temperature
for the polyacrylates shown in figure \ref{cap:Polycarylate-used-as},
with $n_{s}=2$, $n_{t}=30$, $\delta t=2$ and $n_{r}=9$.

This SCLCP has a isotropic-nematic transition at $T_{IN}^{exp}=425K$
\cite{Finkelmann1984}. As a test for the validity of the this correspondence
between temperature and $\chi$ parameter, we calculated the transition
temperature for this polymer using our model. Setting $\frac{\omega}{\chi}=4.6$
and $B=-0.05$ we obtain a critical temperature of $T_{IN}^{teo}=430K$.
Since we have done several approximations in the model, and the results
depend on two adjustable parameters, we must not be too optimistic
about the agreement of the model presented here with respect to the
experimental data. Nevertheless, the agreement achieved can be considered
as a good indicative of the possibility that main physical features
of the systems are present in our phenomenological model.

From now on, we will use relation (\ref{eq:chiflory}) to talk about
temperature instead of the Flory-Huggins parameter. In this way our
discussions will have a more intuitive background.

\subsection{Influence of the structural parameters on the phase diagram morphology}

According to Finkelmann and Rehage \cite{Finkelmann1984}, the schematic
phase behavior of SCLCP follows the same trends as low molar mass
liquid crystals. For a fixed mesogen length, increasing the length
of the flexible spacer or the length of the backbone segment neighboring
side chains decreases the instability temperature towards nematic
or smectic mesophases. Some aspects of our results are however different.
First, in our model, for a fixed $\delta t$, the smectic phase is
favored as $n_{s}$ is increased only if $\omega<\omega_{c}$. Here,
$\omega_{c}$ is a function of the isotropic-microphase transition
temperature curve. It is defined as the curve $\omega_{c}$ that crosses
the curve IM at the point at which it starts decreasing with increasing
$n_{s}$. For larger Maier-Saupe couplings the general trend is opposite
to that related by Finkelmann and Rehage. In figure \ref{cap:Transition-temperatures-as}
we show the phase diagram as a function of the spacer length for two
different values of $\frac{\omega}{\chi}$. In the first case, the
behavior is as described by Finkelmann and Rehage, while in the second
case, where the $\frac{\omega}{\chi}$ ratio is larger, the smectic
phase is not necessarily favored by increasing $n_{s}$ . In figure
\ref{cap:Transition-temperatures-dt5} we show the phase diagrams
for a larger value of $\delta t$. As it can be deduced from the figure,
the behaviors described above are maintained despite the larger value
of $\delta t$.

\begin{figure}
\begin{center}\psfrag{5}{$5$}

\psfrag{4}{$4$}

\psfrag{3}{$3$}

\psfrag{2}{$2$}

\psfrag{1}{$1$}

\psfrag{2}{$2$}

\psfrag{4}{$4$}

\psfrag{6}{$6$}

\psfrag{8}{$8$}

\psfrag{10}{$10$}

\psfrag{12}{$12$}

\psfrag{14}{$14$}

\psfrag{Tcrit}{$T\left( K \right)$}

\psfrag{I-N}{$IN curves$}

\psfrag{I-M}{$IM curve$}

\psfrag{nr}{$n_r$}

\psfrag{w/c=3}{$\frac{\omega}{\chi}=3$}

\psfrag{w/c=4}{$\frac{\omega}{\chi}=4$}

\psfrag{w/c=5}{$\frac{\omega}{\chi}=5$}

\includegraphics[%
  width=0.9\columnwidth,
  keepaspectratio]{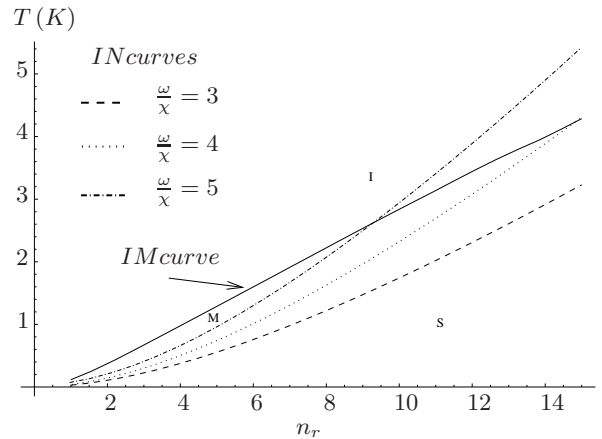}\end{center}

\caption{\label{cap:Tvrsnr}Transition temperatures as a function of the mesogen
length.The parameters were set as $\delta t=3$, $n_{s}=10$, $n_{t}=30$.}
\end{figure}

As discussed by Auriemma\cite{AURIEMMA1990} et al. and by Finkelmann
and Rehage \cite{Finkelmann1984}, another characteristic behavior
of low molecular weight liquid crystals and enhanced in SCLCP, is
the increase of the isotropization temperature with increasing length
of the mesogenic groups. This trend is also present in our model.
In figure \label{cap:Tvrsnr}, we show the transition temperature
as a function of $n_{r}$ for different values of the ratio $\frac{\omega}{\chi}$.
In this case, the smectic phase is favoured by both the increase of
$n_{r}$ the interaction ratio $\frac{\omega}{\chi}$.

\subsection{Microphase segregation}

As mentioned before, an important feature of our model, is the possibility
of microphase segregation besides the formation of nematic and smectic
phases. In this sense, our phase diagrams for SCLCP are more similar
to rod-coil diblock copolymers phase diagrams. In these systems, the
mutual repulsion of the blocks, due to the difference in chain rigidity,
and the constrains imposed by the connectivity of the blocks result
into the formation of supramolecular structures as small as few nanometers\cite{lee2001ssr}.
If we think about the SCLCP as being a polymer in which the monomeric
units are rod-coil diblock copolymers, it is quite natural to expect
the existence of such microphases.

This kind of interplay between liquid crystalline mesophases and microstructure
domains is documented experimentally for SCLCP made of diblock copolymers,
as in references \cite{lee2002mst}\cite{ivanova2004ibd} . In these
SCLCP, the repeated unit is a homogeneous diblock copolymer with a
liquid-crystalline monomer grafted to one of the blocks. Such kind
of SCLCP exhibit microdomains with spherical, hexagonally packed cylindrical
and lamellar structure. The morphology of their phase diagram can
be related to the molecular weight, the block composition and the
$\chi$ parameter. 

Therefore, the existence of microphases in SCLCP is a possibility
that should be considered when performing experiments and analyzing
experimental data.

It is worth to note that despite our model predicts microphase segregation
for SCLCP with certain structural characteristics, we can not differentiate
the symmetries of this microphases. To distinguish between the different
possible symmetries, it would bee necessary to perform calculations
with higher order terms of the free energy, as done by Reenders et
al. \cite{Reenders2002} for the case of rod-coil diblock copolymers.

\subsection{Frank elastic constants}

As explained in the review written by Stephen and Straley\cite{Stephen1974},
the gradient terms in $Q$ of the free energy (\ref{eq:fspin}) are
related to the Frank elastic constants. 

The usual expression for the Frank elastic energy in terms of the
director $\mathbf{n}$ is

\begin{eqnarray}
F_{elastic} & = & \frac{1}{2}K_{1}\left(\nabla\cdot\mathbf{n}\right)^{2}+\frac{1}{2}K_{2}\left(\mathbf{n}\cdot\nabla\times\mathbf{n}\right)^{2}\nonumber \\
 &  & +\frac{1}{2}K_{3}\left(\mathbf{n}\times\nabla\times\mathbf{n}\right)^{2}.\label{eq:frank1}\end{eqnarray}

The constants $K_{i}$ are the splay, twist and bend Frank elastic
constants. In general, the bending constant $K_{3}$ is larger than
the other two, which are about the same order of magnitude. In many
cases equation (\ref{eq:frank1}) is too complex to be of practical
use. Thus, it might be useful consider all the constants as equal.
In that case the elastic energy is reduced to

\begin{equation}
F_{elastic}\approx\frac{1}{2}K\,\partial_{i}n_{j}\partial_{j}n_{i},\label{eq:frank2}\end{equation}
where the sum over repeated indexes is employed.

Equation (\ref{eq:frank2}) is not quantitatively correct but gives
a qualitative idea of the distortions in nematic liquid crystals \cite{degennes1993plc}.

\begin{figure}
\psfrag{35}{$35$}

\psfrag{30}{$30$}

\psfrag{25}{$25$}

\psfrag{20}{$20$}

\psfrag{15}{$15$}

\psfrag{10}{$10$}

\psfrag{5}{$5$}

\psfrag{0.2}{$0.2$}

\psfrag{0.4}{$0.4$}

\psfrag{0.6}{$0.6$}

\psfrag{0.8}{$0.8$}

\psfrag{GQQ}{$G_{QQ} \left( q \right )\times 10^3$}

\psfrag{qR}{$qR$}\includegraphics[%
  width=0.9\columnwidth]{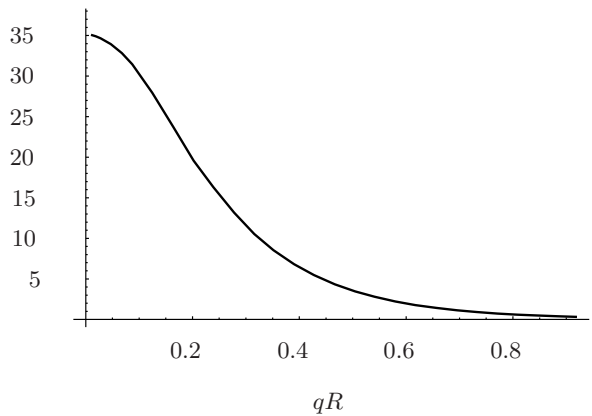}

\caption{\label{cap:GQQ}Correlation function for the nematic tensor calculated
for $\delta t=3$, $n_{s}=7$,$n_{r}=10$, $n_{t}=30,$ $\frac{\omega}{\chi}=4$
and $T=1.1T_{crit}$.}
\end{figure}

In figure \ref{cap:GQQ} we show the correlation function for the
nematic tensor calculated near the isotropic-nematic transition temperature.
This correlation function can be approximated by a function of the
kind

\begin{equation}
G_{QQ}\left(q\right)\approx\frac{1}{\mu+Lq^{2}}.\label{eq:Gqq}\end{equation}

The inverse of $G_{QQ}$ corresponds to the expansion of the free
energy term proportional to the square of the nematic order parameter
in power series of the wave vector. The coefficient $\mu$ carries
the dependence on the Maier-Saupe coupling. When $\mu=0$, the correlation
function diverges at $q=0$. Therefore, $\mu$ is defined by the the
IN curve given by equation (\ref{eq:MS}),

\[
\mu=\frac{5}{f_{r}^{2}}-\frac{2\omega}{3}\frac{N}{n_{t}}.\]

The coefficient $L$ , which corresponds to 

\begin{widetext}

\begin{equation}
L=\frac{n_{t}\left[n_{r}^{2}\left(137+28n_{t}\right)\left(n_{s}+\delta t\right)^{2}+220n_{s}\left(n_{s}^{2}+3n_{s}\delta t+3\delta t^{2}\right)\right]}{7\left[3n_{r}^{2}\left(n_{s}+\delta t\right)^{2}+4n_{s}\left(n_{s}^{2}+3n_{s}\delta t+3\delta t^{2}\right)\right]},\label{eq:K}\end{equation}

\end{widetext}

depends only on the structural parameters $n_{r}$,$n_{s}$,$\delta t$,
and $n_{t}$. Regarding to the form of (\ref{eq:Gqq}) , we can interpret
the quantity $\xi_{n}^{2}=\frac{L}{\mu}$ as a correlation length
related to the nematic ordering. Also, coefficient $L$ is the gradient
term of the free elastic energy and thus can be associated to the
elastic constant $K$. Note that this elasticity is purely entropic.
Therefore, by studying the behavior of the correlation function $G_{QQ}$
with the degree of polymerization and the structural parameter of
the SCLCP, its possible to obtain information of how the polymerization
degree and the size of the spacer affect the nematic ordering.

\begin{figure}
\psfrag{400}{$400$}

\psfrag{300}{$300$}

\psfrag{200}{$200$}

\psfrag{100}{$100$}

\psfrag{10}{$10$}

\psfrag{20}{$20$}

\psfrag{30}{$30$}

\psfrag{40}{$40$}

\psfrag{50}{$50$}

\psfrag{K}{$\frac{K \left( n_{t} \right)}{K \left( 1 \right)}$}

\psfrag{nt}{$n_t$}

\psfrag{ns4}{$n_s=n_r/2$}

\psfrag{ns8}{$n_s=n_r$}

\psfrag{ns16}{$n_s=2n_r$}\includegraphics[%
  width=0.9\columnwidth]{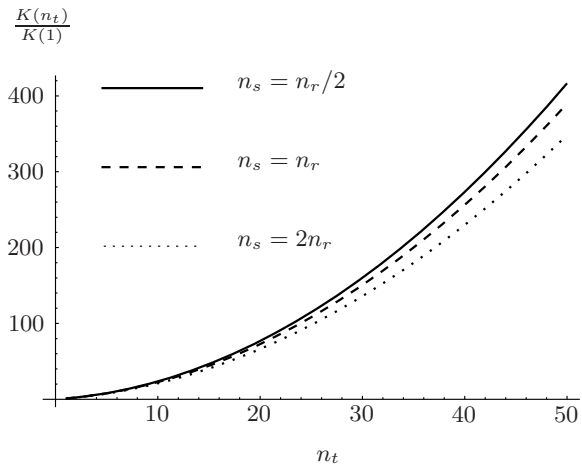}

\caption{\label{cap:Elastic-constant}Elastic constant $K$ as a function
of the polymerization degree for various values of $n_{s}$ at a temperature
$T=1.1T_{crit}$. The other parameters were set as $f_{r}=0.4$, $\delta t=3$,
$\frac{\omega}{\chi}=4,$ $n_{t}=30$.}
\end{figure}

In figure \ref{cap:Elastic-constant} we show the coefficient $K$
as a function of the polymerization degree for various values of $n_{s}$
near the isotropic-nematic transition temperature. Clearly, the additional
correlations introduced polymerization degree raise the energy cost
of the nematic distortions. This reflects as a rise of the nematic
correlation length. From (\ref{eq:K}) it is clear that $K$ has an
asymptotic behavior of $n_{t}^{2}$. Therefore the nematic correlation
length scales as $n_{t}$. While in low molar mass liquid the elastic
energy represents a small fraction of the overall energy \cite{degennes1993plc},
its contribution can be amplified considerably by the polymerization
in SCLCP .

On the other hand, the increase of the spacer size slightly reduces
the value of $K$. The longer the spacer, the weaker the coupling
between the mesogens and backbone conformation and the shortest the
nematic correlation length. 

\subsection{Correlation length for the density fluctuations}

\begin{figure}
\psfrag{G}{$G_{\phi\phi}\left( q \right)\times 10^4$}

\psfrag{0.0175}{$175$}

\psfrag{0.015}{$150$}

\psfrag{0.0125}{$125$}

\psfrag{0.01}{$100$}

\psfrag{0.0075}{$75$}

\psfrag{0.005}{$50$}

\psfrag{0.0025}{$25$}

\psfrag{0.5}{$0.5$}

\psfrag{1}{$1.0$}

\psfrag{1.5}{$1.5$}

\psfrag{2}{$2.0$}

\psfrag{q}{$qR$}

\psfrag{1.43}{$1.43T_{crit}$}

\psfrag{1.25}{$1.25T_{crit}$}

\psfrag{1.1}{$1.1T_{crit}$}\includegraphics[%
  width=0.9\columnwidth]{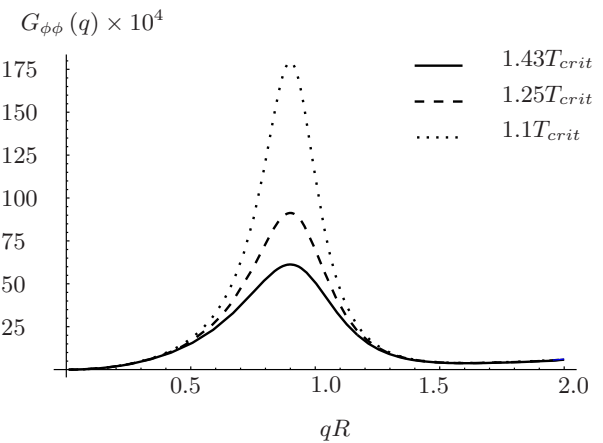}

\caption{\label{cap:Volume-fraction-correlation}Volume fraction correlation
function for temperature close to the transition. The parameters used
are $\delta t=3$, $n_{r}=10$, $n_{s}=12$, and $\frac{\omega}{\chi}=4$
.}
\end{figure}

Information about the correlation length can be extracted from the
volume fraction correlation function in the reciprocal space, $G_{\phi\phi}\left(q\right)$.
This function is calculated as the first element of the inverse of
matrix $\Gamma^{\left(2\right)\mu\nu\rho\sigma}$. In figure \ref{cap:Volume-fraction-correlation}
we show $G_{\phi\phi}\left(q\right)$ for different temperatures.
Even at very high temperatures, its maximum is always at $q^{*}\neq0$,
i.e., the homogeneous state is not the lowest free energy state. Therefore,
we suppose that near the transition temperature, the correlation function
$G_{\phi\phi}\left(q\right)$ is proportional to a function of the
kind

\[
\mathcal{G}_{\phi\phi}\left(q\right)=\frac{1}{r+q^{2}+\frac{u}{q^{2}}}.\]

For temperatures very close to the transition temperature, we can
use the approximation

\[
\mathcal{G}_{\phi\phi}\left(q\right)\approx\frac{1}{\left(q-q_{0}\right)^{2}+\xi^{-2}}\]
where $q_{0}$ is the wave vector that maximizes $\mathcal{G}_{\phi\phi}\left(q\right)$,
and $\xi$ is the correlation length.

In figure we show the correlation length $\xi$ as a function of the
reduced temperature $t=\frac{T-T_{c}}{T_{c}}$ for different degrees
of polymerization.The change in the value of the correlation length
from $n_{t}=1$ to $n_{t}=2$ clearly shows the the importance of
the correlations induced by the linkage of the mesogens to the backbone.
As the polymerization number increases, the correlation length increases.
As discussed by Finkelmann and Rehage \cite{Finkelmann1984}, after
a certain number of monomers added to the SCLCP, the correlation between
the last and the first monomer is lost. 

\begin{figure}
\psfrag{x}{$\xi \left(t \right)q0 $}

\psfrag{t}{$t$}

\psfrag{10}{$10$}

\psfrag{8}{$8$}

\psfrag{6}{$6$}

\psfrag{4}{$4$}

\psfrag{2}{$2$}

\psfrag{0.1}{$0.1$}

\psfrag{0.2}{$0.2$}

\psfrag{0.3}{$0.3$}

\psfrag{0.4}{$0.4$}

\psfrag{nt1}{$n_t=1$}

\psfrag{nt2}{$n_t=2$}

\psfrag{nt3}{$n_t=3$}

\psfrag{nt5}{$n_t=5$}

\psfrag{nt10}{$n_t=15$}

\psfrag{nt30}{$n_t=30$}\includegraphics[%
  width=0.9\columnwidth]{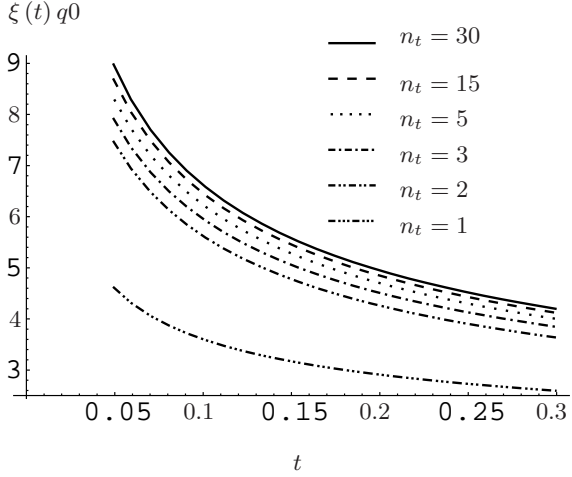}

\caption{\label{cap:Correlation-length-as}Correlation length as a function
of the reduced temperature for different values of the polymerization
degree.The parameters used are $\delta t=3$, $n_{r}=10$, $n_{s}=12$,
$\frac{\omega}{\chi}=4$ and $n_{t}=30$.}
\end{figure}

As expected for a second order transition, the correlation length
diverges at $t=0$. Nevertheless, the form of the correlation function
is such that all the allowed wave vectors form a spherical surface
of radius $q_{0}$. Such a condition is characteristic of systems
with a fluctuation induced first order transition, as deduced by Brazovskii\cite{Brazovskii1975}.
Higher order calculations would be necessary to corroborate this statement.

\section{Conclusions}

As discussed above, the model we have proposed for SCLCP is in good
agreement with the experimental observed trends discussed by Finkelmann
and Rehage \cite{Finkelmann1984} and Shibaev and Platé\cite{Shibaev1984}
for the transition temperatures of the SCLCP. Moreover, for and adequate
value of the ratio $\frac{\omega}{\chi}$, our model predicts the
existence of microphases with a modulated volume fraction but without
orientational order . These microphases, originated from the mutual
repulsion between the rigid and flexible blocks of the SCLCP, are
already observed in SCLCP made of diblock copolymers as explained
in references \cite{lee2002mst}\cite{ivanova2004ibd}. Nevertheless,
further calculations are necessary in order to identify their symmetry
.

As expected, the additional correlations introduced by the linkage
of the mesogenic groups to a polymeric backbone through the flexible
have a strong influence on the SCLCP critical behavior. This statement
is supported by two main features. The first one is the swelling of
the characteristic length of the microphase domains, that we have
attributed to the backbone conformation. The second one is the the
influence of the spacer length $n_{s}$, and polymerization degree
$n_{t}$ on the phase diagram topology, correlation length of the
volume fraction correlations and elastic energy associated to the
nematic distortions.

\begin{acknowledgments}
M.H. thanks to FAPESP, project 03/06370-2, for financial support 

H.W. thanks FAPESP and CNPq for financial support.
\end{acknowledgments}
\appendix

\section{The noninteracting correlation functions\label{sec:The-non-interacting}}

The correlation functions are calculated as explained in references
\cite{Reenders2002,Shinozaki1994}, with the difference that there
is a rigid rod attached at the end of each tooth of the comb polymer. 

As an example, we calculate the correlation function rod-rod. This
function has two contributions: one that comes from the correlation
of the segments within the rod and the other one coming form the interaction
between segments of two different rods. In this way:

\begin{widetext}\begin{eqnarray}
G_{rr}\left(\mathbf{q}\right) & = & \frac{1}{N^{2}}n_{t}\int_{0}^{N_{r}}d\tau\int_{0}^{N_{r}}d\tau'\left\langle e^{i\left(\tau-\tau'\right)b\mathbf{u}.\mathbf{q}}\right\rangle \nonumber \\
 &  & +\frac{1}{N^{2}}e^{-2\left(qR_{s}\right)^{2}}\sum_{i\neq j}^{N_{t}}e^{-\frac{\left(qb\right)^{2}}{6}\delta t\left|i-j\right|}\int_{0}^{N_{r}}d\tau\int_{0}^{N_{r}}d\tau'\left\langle e^{i\tau b\mathbf{u}^{i}.\mathbf{q}+\tau'b\mathbf{u}^{j}.\mathbf{q}}\right\rangle \nonumber \\
 & = & f_{r}^{2}\left[\frac{1}{n_{t}}K_{rr}\left(N_{r}bq\right)+e^{-2q^{2}R_{s}^{2}}F_{R}\left(N_{r}bq\right)^{2}\left(D_{n_{t}}\left(qR_{b}\right)-\frac{1}{n_{t}}\right)\right].\label{eq:GRR}\end{eqnarray}

\end{widetext}where, $K_{rr}\left(x\right)$, $F_{r}\left(x\right)$
and $D_{n}\left(x\right)$ are the the rod-rod correlation function,
thee form factor of a rod of length $l$ and the Debye function for
the backbone as explained in the text.

The other correlation function, calculated in the same way, are

\[
G_{bb}\left(\mathbf{q}\right)=f_{b}^{2}D_{n_{b}}\left(qR_{b}\right),\]

\begin{eqnarray*}
G_{ss}\left(\mathbf{q}\right) & = & \frac{f_{s}^{2}}{n_{t}}D_{n_{s}}\left(qR_{s}\right)+\\
 &  & f_{s}^{2}F_{n_{s}}\left(qR_{s}\right)^{2}\left(D_{n_{s}}\left(qR_{b}\right)-\frac{1}{n_{t}}\right),\end{eqnarray*}
where $F_{n_{s}}\left(qR_{s}\right)=\frac{1}{n_{s}}\left(\frac{1-e^{-q^{2}R_{s}^{2}}}{e^{\frac{q^{2}R_{s}^{2}}{n_{s}}}-1}\right)$
is the structure factor of the spacer,

\begin{eqnarray*}
G_{rs}\left(\mathbf{q}\right) & = & f_{r}f_{s}F_{r}\left(qbn_{r}\right)F_{n_{s}}\left(qR_{s}\right)\\
 &  & \times\left[\frac{1}{n_{t}}+\left(D_{n_{t}}\left(qR_{b}\right)-\frac{1}{n_{t}}\right)e^{-\left(qR_{s}\right)^{2}}\right]\end{eqnarray*}

\begin{eqnarray*}
G_{rb}\left(\mathbf{q}\right) & = & 2f_{r}f_{b}\frac{6}{b^{2}q^{2}}F_{r}\left(qbn_{r}\right)e^{-\left(qR_{s}\right)^{2}}\\
 &  & \times\left[1-e^{\frac{1}{12}b^{2}q^{2}\delta t}F_{n_{t}}\left(qR_{b}\right)\right],\end{eqnarray*}
where $F_{n_{t}}\left(qR_{b}\right)=\frac{1}{n_{b}}\left(\frac{1-e^{-q^{2}R_{b}^{2}}}{e^{\frac{q^{2}R_{b}^{2}}{n_{b}}}-1}\right)$
is the structure factor of the backbone segments between two side
groups,

\[
G_{sb}\left(\mathbf{q}\right)=2cf_{b}f_{s}F_{n_{s}}\left(qR_{s}\right)\frac{6}{b^{2}q^{2}n_{b}}\left[1-e^{\frac{1}{12}b^{2}q^{2}\delta t}F_{n_{t}}\left(qR_{b}\right)\right],\]
where we have defined $\Delta^{\mu\nu}\equiv\frac{q^{\mu}q^{\nu}}{q^{2}}-\frac{\delta^{\mu\nu}}{3}$,

\begin{eqnarray*}
G_{Qs}^{\mu\nu}\left(\mathbf{q}\right) & = & f_{r}f_{s}\mathbf{\Delta}^{\mu\nu}F_{n_{s}}\left(qR_{s}\right)K_{Q_{r}r}\left(n_{r}bq\right)\\
 &  & \times\left\{ \frac{1}{n_{t}}+e^{-\left(qR_{s}\right)^{2}}\left[D_{n_{t}}\left(qR_{b}\right)-\frac{1}{n_{t}}\right]\right\} ,\end{eqnarray*}

\begin{eqnarray*}
G_{Qb}^{\mu\nu}\left(\mathbf{q}\right) & = & f_{r}f_{b}F_{Q_{r}}\left(ql\right)e^{-\left(qR_{s}\right)^{2}}\frac{6}{b^{2}q^{2}n_{b}}\\
 & \times & \left[1-e^{\frac{1}{12}q^{2}b^{2}\delta t}F_{n_{t}}\left(qR_{b}\right)\right]\end{eqnarray*}
where $F_{Q_{r}}\left(x\right)=\frac{1}{2x^{3}}\left[3\sin\left(x\right)-3x\cos\left(x\right)-x^{2}Si\left(x\right)\right]$
and $K_{Qr}\left(x\right)=\frac{4x-x\cos x-3\sin x-x^{2}Si\left(x\right)}{x^{3}}$
as defined in reference \cite{Reenders2002},

\begin{eqnarray*}
G_{Qr}^{\mu\nu}\left(\mathbf{q}\right) & = & f_{r}^{2}\mathbf{\Delta}_{\mu\nu}\left[\frac{1}{n_{t}}K_{Q_{r}r}\left(ql\right)+e^{-2\left(qR_{s}\right)^{2}}\right.\\
 &  & \left.\times F_{n_{r}}\left(ql\right)F_{Q_{r}}\left(ql\right)\left(D_{n_{t}}\left(qR_{b}\right)-\frac{1}{n_{t}}\right)\right]\end{eqnarray*}

\begin{eqnarray*}
G_{QQ}^{\mu\nu\rho\sigma}\left(\mathbf{q}\right) & = & f_{r}^{2}\sum_{i=1}^{3}\mathcal{T}_{i}^{\mu\nu,\rho\sigma}\left[\frac{1}{n_{t}}K_{Si}^{\left(2\right)}\right.\\
 &  & \left.+e^{-2\left(qR_{s}\right)^{2}}F_{Q_{r}}\left(ql\right)^{2}\left(D_{n_{t}}\left(qR_{b}\right)-\frac{1}{n_{t}}\right)\delta_{i,3}\right],\end{eqnarray*}
where the tensors $\mathcal{T}_{i}^{\mu\nu,\rho\sigma}$ and the functions
$K_{si}$ are defined as in reference \cite{Reenders2002}.

Therefore, the correlations used to calculate matrix \ref{eq:correlation matrix.}
are defined as follows:

\[
G_{cr}\left(\mathbf{q}\right)=G_{rs}\left(\mathbf{q}\right)+G_{rb}\left(\mathbf{q}\right),\]

\[
G_{cc}\left(\mathbf{q}\right)=G_{ss}\left(\mathbf{q}\right)+G_{bb}\left(\mathbf{q}\right)+2G_{bs}\left(\mathbf{q}\right),\]
and

\[
G_{Qc}\left(\mathbf{q}\right)=G_{Qs}+G_{Qb}.\]

\section{The Hildebrand parameters\label{sec:The-Hildebrand-parameters}}

According to reference \cite{grulke1998ph}, for non polar substances,
the solubility parameter can be calculated through its relationship
with the cohesive energy and its molar volume:

\begin{equation}
\delta=\sqrt{\frac{E_{coh}}{V}}\label{eq:Hnopolar}\end{equation}

Using to the group contribution method, the solubility parameter of
a monomer unit is expressed as the sum of independent contributions
from the atomic groups that constitute the unit,

\begin{equation}
E_{coh}=\sum_{i}E_{coh}^{\left(i\right)}.\label{eq:Ecoh}\end{equation}

In the case of interest, we are considering the mixture of two components:
the comb polymer (backbone and spacer). Therefore, it is convenient
to choose the the segment unit as having a molar volume equivalent
to that of the CH$_{2}$. In this way, using (\ref{eq:Hnopolar})
and the with the data in reference is \cite{grulke1998ph}, we find
that the Hildebrant solubility parameter for the flexible part of
the polyacrylate is

\[
\delta_{pol}=16.13\left(Jcm^{-3}\right)^{1/2}.\]

For the rigid part, we must divide the cohesion energy of the whole
group by $n_{r}$. This is equivalent to substitute the rigid rod
by a chain of $n_{r}$ units of $CH_{2}$ each one with a solubility
parameter $\delta_{solv}=\sqrt{\frac{E_{rod}^{coh}}{n_{r}V}}$. In
this way we obtain the Hildebrant solubility parameter for the rigid
core,\[
\delta_{solv}=7.73\left(Jcm^{-3}\right)^{1/2}.\]

By substituting and in $\delta_{pol}$ and $\delta_{solv}$ in \ref{eq:chi-hildebrand},
it is possible to calculated transition temperatures.

\def\urlprefix{}    

\def\url#1{}

\bibliographystyle{apsrev}
\bibliography{comb-lc_c}

\begin{thebibliography}{38}
\expandafter\ifx\csname natexlab\endcsname\relax\def\natexlab#1{#1}\fi
\expandafter\ifx\csname bibnamefont\endcsname\relax
  \def\bibnamefont#1{#1}\fi
\expandafter\ifx\csname bibfnamefont\endcsname\relax
  \def\bibfnamefont#1{#1}\fi
\expandafter\ifx\csname citenamefont\endcsname\relax
  \def\citenamefont#1{#1}\fi
\expandafter\ifx\csname url\endcsname\relax
  \def\url#1{\texttt{#1}}\fi
\expandafter\ifx\csname urlprefix\endcsname\relax\def\urlprefix{URL }\fi
\providecommand{\bibinfo}[2]{#2}
\providecommand{\eprint}[2][]{\url{#2}}

\bibitem[{\citenamefont{Shibaev and Platé}(1984)}]{Shibaev1984}
\bibinfo{author}{\bibfnamefont{V.~P.} \bibnamefont{Shibaev}} \bibnamefont{and}
  \bibinfo{author}{\bibfnamefont{N.~A.} \bibnamefont{Platé}},
  \bibinfo{journal}{Advances in Polymer Science} \textbf{\bibinfo{volume}{60}},
  \bibinfo{pages}{173} (\bibinfo{year}{1984}).

\bibitem[{\citenamefont{Lee et~al.}(2001)\citenamefont{Lee, Cho, and
  Zin}}]{lee2001ssr}
\bibinfo{author}{\bibfnamefont{M.}~\bibnamefont{Lee}},
  \bibinfo{author}{\bibfnamefont{B.}~\bibnamefont{Cho}}, \bibnamefont{and}
  \bibinfo{author}{\bibfnamefont{W.}~\bibnamefont{Zin}},
  \bibinfo{journal}{Chemical Review} \textbf{\bibinfo{volume}{101}},
  \bibinfo{pages}{3869} (\bibinfo{year}{2001}).

\bibitem[{\citenamefont{Yamazaki et~al.}(2004)\citenamefont{Yamazaki, Motoyama,
  Nonomura, and Ohta}}]{Yamazaki2004}
\bibinfo{author}{\bibfnamefont{N.}~\bibnamefont{Yamazaki}},
  \bibinfo{author}{\bibfnamefont{M.}~\bibnamefont{Motoyama}},
  \bibinfo{author}{\bibfnamefont{M.}~\bibnamefont{Nonomura}}, \bibnamefont{and}
  \bibinfo{author}{\bibfnamefont{T.}~\bibnamefont{Ohta}}, \bibinfo{journal}{The
  Journal of Chemical Physics} \textbf{\bibinfo{volume}{120}},
  \bibinfo{pages}{3949} (\bibinfo{year}{2004}).

\bibitem[{\citenamefont{Pryamitsyn and Ganesan}(2004)}]{Pryamitsyn2004}
\bibinfo{author}{\bibfnamefont{V.}~\bibnamefont{Pryamitsyn}} \bibnamefont{and}
  \bibinfo{author}{\bibfnamefont{V.}~\bibnamefont{Ganesan}},
  \bibinfo{journal}{The Journal of Chemical Physics}
  \textbf{\bibinfo{volume}{120}}, \bibinfo{pages}{5824} (\bibinfo{year}{2004}).

\bibitem[{\citenamefont{Renz and Warner}(1986)}]{Renz1986}
\bibinfo{author}{\bibfnamefont{W.}~\bibnamefont{Renz}} \bibnamefont{and}
  \bibinfo{author}{\bibfnamefont{M.}~\bibnamefont{Warner}},
  \bibinfo{journal}{Physical Review Letters} \textbf{\bibinfo{volume}{56}},
  \bibinfo{pages}{1268} (\bibinfo{year}{1986}).

\bibitem[{\citenamefont{Finkelmann}(1991)}]{finkelmann1991cel}
\bibinfo{author}{\bibfnamefont{H.}~\bibnamefont{Finkelmann}},
  \emph{\bibinfo{title}{Liquid crystallinity in Polymers}}
  (\bibinfo{publisher}{New York: VCH Publishers Inc}, \bibinfo{year}{1991}).

\bibitem[{\citenamefont{Reenders and ten Brinke}(2002)}]{Reenders2002}
\bibinfo{author}{\bibfnamefont{M.}~\bibnamefont{Reenders}} \bibnamefont{and}
  \bibinfo{author}{\bibfnamefont{G.}~\bibnamefont{ten Brinke}},
  \bibinfo{journal}{Macromolecules} \textbf{\bibinfo{volume}{35}},
  \bibinfo{pages}{3266} (\bibinfo{year}{2002}).

\bibitem[{\citenamefont{Motoyama et~al.}(2003)\citenamefont{Motoyama, Yamazaki,
  Nonomura, and Ohta}}]{Motoyama2003}
\bibinfo{author}{\bibfnamefont{M.}~\bibnamefont{Motoyama}},
  \bibinfo{author}{\bibfnamefont{N.}~\bibnamefont{Yamazaki}},
  \bibinfo{author}{\bibfnamefont{M.}~\bibnamefont{Nonomura}}, \bibnamefont{and}
  \bibinfo{author}{\bibfnamefont{T.}~\bibnamefont{Ohta}},
  \bibinfo{journal}{Jounal of the Physical Society of Japan}
  \textbf{\bibinfo{volume}{72}}, \bibinfo{pages}{991} (\bibinfo{year}{2003}).

\bibitem[{\citenamefont{Matsen and Barrett}(1998)}]{matsen:4108}
\bibinfo{author}{\bibfnamefont{M.~W.} \bibnamefont{Matsen}} \bibnamefont{and}
  \bibinfo{author}{\bibfnamefont{C.}~\bibnamefont{Barrett}},
  \bibinfo{journal}{The Journal of Chemical Physics}
  \textbf{\bibinfo{volume}{109}}, \bibinfo{pages}{4108} (\bibinfo{year}{1998}),
  \urlprefix\url{http://link.aip.org/link/?JCP/109/4108/1}.

\bibitem[{\citenamefont{Shinozaki et~al.}(1994)\citenamefont{Shinozaki, Jasnow,
  and Balazs}}]{Shinozaki1994}
\bibinfo{author}{\bibfnamefont{A.}~\bibnamefont{Shinozaki}},
  \bibinfo{author}{\bibfnamefont{D.}~\bibnamefont{Jasnow}}, \bibnamefont{and}
  \bibinfo{author}{\bibfnamefont{A.~C.} \bibnamefont{Balazs}},
  \bibinfo{journal}{Macromolecules} \textbf{\bibinfo{volume}{27}},
  \bibinfo{pages}{2496 } (\bibinfo{year}{1994}).

\bibitem[{\citenamefont{Vlahos and Kosmas}(1987)}]{Vlahos1987}
\bibinfo{author}{\bibfnamefont{C.~H.} \bibnamefont{Vlahos}} \bibnamefont{and}
  \bibinfo{author}{\bibfnamefont{M.~K.} \bibnamefont{Kosmas}},
  \bibinfo{journal}{J. Phys. A: Math. Gen.} \textbf{\bibinfo{volume}{20}},
  \bibinfo{pages}{1471} (\bibinfo{year}{1987}).

\bibitem[{\citenamefont{Wang et~al.}(2005)\citenamefont{Wang, Jiang, and
  Hu}}]{Wang2005}
\bibinfo{author}{\bibfnamefont{R.}~\bibnamefont{Wang}},
  \bibinfo{author}{\bibfnamefont{Z.}~\bibnamefont{Jiang}}, \bibnamefont{and}
  \bibinfo{author}{\bibfnamefont{J.}~\bibnamefont{Hu}},
  \bibinfo{journal}{Polymer} \textbf{\bibinfo{volume}{46}},
  \bibinfo{pages}{6201} (\bibinfo{year}{2005}),
  \urlprefix\url{http://www.sciencedirect.com/science/article/B6TXW-4GDBTH1-9/%
2/088d1f7f52f4bba22d2a472be635da38}.

\bibitem[{\citenamefont{Sergei V.~Vasilenko}(1985)}]{Sergei1985a}
\bibinfo{author}{\bibfnamefont{A.~R.~K.} \bibnamefont{Sergei V.~Vasilenko},
  \bibfnamefont{Valery P.~Shibaev}}, \bibinfo{journal}{Die Makromolekulare
  Chemie} \textbf{\bibinfo{volume}{186}}, \bibinfo{pages}{1951}
  (\bibinfo{year}{1985}),
  \urlprefix\url{http://dx.doi.org/10.1002/macp.1985.021860922}.

\bibitem[{\citenamefont{Matheson~Jr and Flory}(1981)}]{mathesonjr1981stm}
\bibinfo{author}{\bibfnamefont{R.}~\bibnamefont{Matheson~Jr}} \bibnamefont{and}
  \bibinfo{author}{\bibfnamefont{P.}~\bibnamefont{Flory}},
  \bibinfo{journal}{Macromolecules} \textbf{\bibinfo{volume}{14}},
  \bibinfo{pages}{954} (\bibinfo{year}{1981}).

\bibitem[{\citenamefont{Auriemma et~al.}(1990)\citenamefont{Auriemma,
  Corradini, and Vacatello}}]{AURIEMMA1990}
\bibinfo{author}{\bibfnamefont{F.}~\bibnamefont{Auriemma}},
  \bibinfo{author}{\bibfnamefont{P.}~\bibnamefont{Corradini}},
  \bibnamefont{and}
  \bibinfo{author}{\bibfnamefont{M.}~\bibnamefont{Vacatello}},
  \bibinfo{journal}{The Journal of Chemical Physics}
  \textbf{\bibinfo{volume}{93}}, \bibinfo{pages}{8314} (\bibinfo{year}{1990}).

\bibitem[{\citenamefont{Wang and Warner}(1987)}]{Wang1987}
\bibinfo{author}{\bibfnamefont{X.~J.} \bibnamefont{Wang}} \bibnamefont{and}
  \bibinfo{author}{\bibfnamefont{M.}~\bibnamefont{Warner}},
  \bibinfo{journal}{J. Phys. A: Math. Gen.} \textbf{\bibinfo{volume}{20}},
  \bibinfo{pages}{713} (\bibinfo{year}{1987}).

\bibitem[{\citenamefont{ten Bosch et~al.}(1983)\citenamefont{ten Bosch, Maissa,
  and Sixou}}]{Bosch1983}
\bibinfo{author}{\bibfnamefont{A.}~\bibnamefont{ten Bosch}},
  \bibinfo{author}{\bibfnamefont{P.}~\bibnamefont{Maissa}}, \bibnamefont{and}
  \bibinfo{author}{\bibfnamefont{P.}~\bibnamefont{Sixou}},
  \bibinfo{journal}{The Journal of Chemical Physics}
  \textbf{\bibinfo{volume}{79}}, \bibinfo{pages}{3462} (\bibinfo{year}{1983}),
  \urlprefix\url{http://link.aip.org/link/?JCP/79/3462/1}.

\bibitem[{\citenamefont{Holyst and Schick}(1992)}]{Holyst1992}
\bibinfo{author}{\bibfnamefont{R.}~\bibnamefont{Holyst}} \bibnamefont{and}
  \bibinfo{author}{\bibfnamefont{M.}~\bibnamefont{Schick}},
  \bibinfo{journal}{The Journal of Chemical Physics}
  \textbf{\bibinfo{volume}{96}}, \bibinfo{pages}{730} (\bibinfo{year}{1992}).

\bibitem[{\citenamefont{Maier and Saupe}(1960)}]{maier1960sms}
\bibinfo{author}{\bibfnamefont{W.}~\bibnamefont{Maier}} \bibnamefont{and}
  \bibinfo{author}{\bibfnamefont{A.}~\bibnamefont{Saupe}}, \bibinfo{journal}{Z.
  Naturforsch, Teil A} \textbf{\bibinfo{volume}{15}}, \bibinfo{pages}{287}
  (\bibinfo{year}{1960}).

\bibitem[{\citenamefont{Stephen and Straley}(1974)}]{Stephen1974}
\bibinfo{author}{\bibfnamefont{M.~J.} \bibnamefont{Stephen}} \bibnamefont{and}
  \bibinfo{author}{\bibfnamefont{J.~P.} \bibnamefont{Straley}},
  \bibinfo{journal}{Reviews of Modern Physics} \textbf{\bibinfo{volume}{46}},
  \bibinfo{pages}{617} (\bibinfo{year}{1974}),
  \urlprefix\url{http://link.aps.org/abstract/RMP/v46/p617}.

\bibitem[{\citenamefont{Gupta and Edwards}(1993)}]{Gupta1993}
\bibinfo{author}{\bibfnamefont{A.~M.} \bibnamefont{Gupta}} \bibnamefont{and}
  \bibinfo{author}{\bibfnamefont{S.~F.} \bibnamefont{Edwards}},
  \bibinfo{journal}{The Journal of Chemical Physics}
  \textbf{\bibinfo{volume}{98}}, \bibinfo{pages}{1588} (\bibinfo{year}{1993}),
  \urlprefix\url{http://link.aip.org/link/?JCP/98/1588/1}.

\bibitem[{\citenamefont{Liu and Fredrickson}(1993)}]{Liu1993}
\bibinfo{author}{\bibfnamefont{A.~J.} \bibnamefont{Liu}} \bibnamefont{and}
  \bibinfo{author}{\bibfnamefont{G.~H.} \bibnamefont{Fredrickson}},
  \bibinfo{journal}{Macromolecules} \textbf{\bibinfo{volume}{26}},
  \bibinfo{pages}{2817} (\bibinfo{year}{1993}).

\bibitem[{\citenamefont{Teraoka}(2002)}]{teraoka2002ps}
\bibinfo{author}{\bibfnamefont{I.}~\bibnamefont{Teraoka}},
  \emph{\bibinfo{title}{Polymer solutions}} (\bibinfo{publisher}{Wiley New
  York}, \bibinfo{year}{2002}).

\bibitem[{\citenamefont{Singh et~al.}(1994)\citenamefont{Singh, Goulian, Liu,
  and Fredrickson}}]{Singh1994}
\bibinfo{author}{\bibfnamefont{C.}~\bibnamefont{Singh}},
  \bibinfo{author}{\bibfnamefont{M.}~\bibnamefont{Goulian}},
  \bibinfo{author}{\bibfnamefont{A.~J.} \bibnamefont{Liu}}, \bibnamefont{and}
  \bibinfo{author}{\bibfnamefont{G.~H.} \bibnamefont{Fredrickson}},
  \bibinfo{journal}{Macromolecules} \textbf{\bibinfo{volume}{27}},
  \bibinfo{pages}{2974} (\bibinfo{year}{1994}).

\bibitem[{\citenamefont{Vilgis}(2000)}]{Vilgis2000}
\bibinfo{author}{\bibfnamefont{T.~A.} \bibnamefont{Vilgis}},
  \bibinfo{journal}{Physics Reports} \textbf{\bibinfo{volume}{336}},
  \bibinfo{pages}{167} (\bibinfo{year}{2000}).

\bibitem[{\citenamefont{de~Gennes}(1979)}]{Gennes1979}
\bibinfo{author}{\bibfnamefont{P.~G.} \bibnamefont{de~Gennes}},
  \emph{\bibinfo{title}{Scaling Concepts in Polymer Physics}}
  (\bibinfo{publisher}{Cornell University Press Ithaca and London},
  \bibinfo{year}{1979}),
  \urlprefix\url{file:/home/westfahl/Documents/ebooks/deGennes1979.djvu}.

\bibitem[{\citenamefont{Finkelmann and Rehage}(1984)}]{Finkelmann1984}
\bibinfo{author}{\bibfnamefont{H.}~\bibnamefont{Finkelmann}} \bibnamefont{and}
  \bibinfo{author}{\bibfnamefont{G.}~\bibnamefont{Rehage}},
  \bibinfo{journal}{Advances in Polymer Science} \textbf{\bibinfo{volume}{60}},
  \bibinfo{pages}{99} (\bibinfo{year}{1984}).

\bibitem[{\citenamefont{Noirez et~al.}(1998)\citenamefont{Noirez, Boeffel, and
  Daoud-Aladine}}]{Noirez1998}
\bibinfo{author}{\bibfnamefont{L.}~\bibnamefont{Noirez}},
  \bibinfo{author}{\bibfnamefont{C.}~\bibnamefont{Boeffel}}, \bibnamefont{and}
  \bibinfo{author}{\bibfnamefont{A.}~\bibnamefont{Daoud-Aladine}},
  \bibinfo{journal}{Physical Review Letters} \textbf{\bibinfo{volume}{80}},
  \bibinfo{pages}{1453} (\bibinfo{year}{1998}),
  \urlprefix\url{http://link.aps.org/abstract/PRL/v80/p1453}.

\bibitem[{\citenamefont{Noirez et~al.}(1995)\citenamefont{Noirez, Keller, and
  Cotton}}]{noirez1995sac}
\bibinfo{author}{\bibfnamefont{L.}~\bibnamefont{Noirez}},
  \bibinfo{author}{\bibfnamefont{P.}~\bibnamefont{Keller}}, \bibnamefont{and}
  \bibinfo{author}{\bibfnamefont{J.}~\bibnamefont{Cotton}},
  \bibinfo{journal}{Liquid Crystals} \textbf{\bibinfo{volume}{18}},
  \bibinfo{pages}{129} (\bibinfo{year}{1995}).

\bibitem[{\citenamefont{Noirez et~al.}(1994)\citenamefont{Noirez, Davidson,
  Schwarz, and Pepy}}]{noirez1994scl}
\bibinfo{author}{\bibfnamefont{L.}~\bibnamefont{Noirez}},
  \bibinfo{author}{\bibfnamefont{P.}~\bibnamefont{Davidson}},
  \bibinfo{author}{\bibfnamefont{W.}~\bibnamefont{Schwarz}}, \bibnamefont{and}
  \bibinfo{author}{\bibfnamefont{G.}~\bibnamefont{Pepy}},
  \bibinfo{journal}{Liquid Crystals} \textbf{\bibinfo{volume}{16}},
  \bibinfo{pages}{1081} (\bibinfo{year}{1994}).

\bibitem[{\citenamefont{Grulke et~al.}(1998)\citenamefont{Grulke, Brandrup, and
  Immergut}}]{grulke1998ph}
\bibinfo{author}{\bibfnamefont{E.}~\bibnamefont{Grulke}},
  \bibinfo{author}{\bibfnamefont{J.}~\bibnamefont{Brandrup}}, \bibnamefont{and}
  \bibinfo{author}{\bibfnamefont{E.}~\bibnamefont{Immergut}},
  \emph{\bibinfo{title}{Polymer handbook}} (\bibinfo{publisher}{Wiley
  Interscience Publication , New York}, \bibinfo{year}{1998}).

\bibitem[{\citenamefont{Almdal et~al.}(2002)\citenamefont{Almdal, Hillmyer, and
  Bates}}]{almdal2002ica}
\bibinfo{author}{\bibfnamefont{K.}~\bibnamefont{Almdal}},
  \bibinfo{author}{\bibfnamefont{M.}~\bibnamefont{Hillmyer}}, \bibnamefont{and}
  \bibinfo{author}{\bibfnamefont{F.}~\bibnamefont{Bates}},
  \bibinfo{journal}{Macromolecules} \textbf{\bibinfo{volume}{35}},
  \bibinfo{pages}{7685} (\bibinfo{year}{2002}).

\bibitem[{\citenamefont{Fredrickson et~al.}(1994)\citenamefont{Fredrickson,
  Liu, and Bates}}]{fredrickson1994ecf}
\bibinfo{author}{\bibfnamefont{G.}~\bibnamefont{Fredrickson}},
  \bibinfo{author}{\bibfnamefont{A.}~\bibnamefont{Liu}}, \bibnamefont{and}
  \bibinfo{author}{\bibfnamefont{F.}~\bibnamefont{Bates}},
  \bibinfo{journal}{Macromolecules} \textbf{\bibinfo{volume}{27}},
  \bibinfo{pages}{2503} (\bibinfo{year}{1994}).

\bibitem[{\citenamefont{Liu and Fredrickson}(1992)}]{liu1992inf}
\bibinfo{author}{\bibfnamefont{A.}~\bibnamefont{Liu}} \bibnamefont{and}
  \bibinfo{author}{\bibfnamefont{G.}~\bibnamefont{Fredrickson}},
  \bibinfo{journal}{Macromolecules} \textbf{\bibinfo{volume}{25}},
  \bibinfo{pages}{5551} (\bibinfo{year}{1992}).

\bibitem[{\citenamefont{LEE and HAN}(2002)}]{lee2002mst}
\bibinfo{author}{\bibfnamefont{K.}~\bibnamefont{LEE}} \bibnamefont{and}
  \bibinfo{author}{\bibfnamefont{C.}~\bibnamefont{HAN}},
  \bibinfo{journal}{Macromolecules} \textbf{\bibinfo{volume}{35}},
  \bibinfo{pages}{3145} (\bibinfo{year}{2002}).

\bibitem[{\citenamefont{Ivanova et~al.}(2004)\citenamefont{Ivanova, Staneva,
  Geppert, Heck, Walter, Gronski, and St{\"u}hn}}]{ivanova2004ibd}
\bibinfo{author}{\bibfnamefont{R.}~\bibnamefont{Ivanova}},
  \bibinfo{author}{\bibfnamefont{R.}~\bibnamefont{Staneva}},
  \bibinfo{author}{\bibfnamefont{S.}~\bibnamefont{Geppert}},
  \bibinfo{author}{\bibfnamefont{B.}~\bibnamefont{Heck}},
  \bibinfo{author}{\bibfnamefont{B.}~\bibnamefont{Walter}},
  \bibinfo{author}{\bibfnamefont{W.}~\bibnamefont{Gronski}}, \bibnamefont{and}
  \bibinfo{author}{\bibfnamefont{B.}~\bibnamefont{St{\"u}hn}},
  \bibinfo{journal}{Colloid \& Polymer Science} \textbf{\bibinfo{volume}{282}},
  \bibinfo{pages}{810} (\bibinfo{year}{2004}).

\bibitem[{\citenamefont{de~Gennes and Prost}(1993)}]{degennes1993plc}
\bibinfo{author}{\bibfnamefont{P.~G.} \bibnamefont{de~Gennes}}
  \bibnamefont{and} \bibinfo{author}{\bibfnamefont{J.}~\bibnamefont{Prost}},
  \emph{\bibinfo{title}{The Physics of Liquid Crystals}}
  (\bibinfo{publisher}{Clarendon, Oxford}, \bibinfo{year}{1993}).

\bibitem[{\citenamefont{Brazovskii}(1975)}]{Brazovskii1975}
\bibinfo{author}{\bibfnamefont{S.~A.} \bibnamefont{Brazovskii}},
  \bibinfo{journal}{Zh. Eksp. Teor. Fiz} \textbf{\bibinfo{volume}{68}},
  \bibinfo{pages}{175} (\bibinfo{year}{1975}).

\end{thebibliography}

\end{document}